\documentclass[twocolumn,floatfix]{aastex63}
\usepackage{apjfonts}
\usepackage{xspace}
\usepackage{color, colortbl}
\definecolor{Gray}{gray}{0.9}
\newcolumntype{g}{>{\columncolor{Gray}[5pt]}c}

\defcitealias{EHT1}{EHT1}
\defcitealias{EHT2}{EHT2}
\defcitealias{EHT3}{EHT3}
\defcitealias{EHT4}{EHT4}
\defcitealias{EHT5}{EHT5}
\defcitealias{EHT6}{EHT6}
\def\sgra{\object{Sgr~A$^{\ast}$}\xspace}
\def\m87{\object{M87$^{\ast}$}\xspace}

\begin{document}

\title{Evaluation of New Submillimeter VLBI Sites for the Event Horizon Telescope}
\author[0000-0002-5779-4767]{Alexander W. Raymond}
\affiliation{Center for Astrophysics $|$ Harvard \& Smithsonian,  60 Garden Street, Cambridge, MA 02138, USA}

\author[0000-0002-7179-3816]{Daniel Palumbo}
\affiliation{Center for Astrophysics $|$ Harvard \& Smithsonian,  60 Garden Street, Cambridge, MA 02138, USA}
\affiliation{Black Hole Initiative, Harvard University, 20 Garden Street, Cambridge, MA 02138, USA}

\author[0000-0003-4622-5857]{Scott N. Paine}
\affiliation{Center for Astrophysics $|$ Harvard \& Smithsonian,  60 Garden Street, Cambridge, MA 02138, USA}

\author[0000-0002-9030-642X]{Lindy Blackburn}
\affiliation{Center for Astrophysics $|$ Harvard \& Smithsonian,  60 Garden Street, Cambridge, MA 02138, USA}
\affiliation{Black Hole Initiative, Harvard University, 20 Garden Street, Cambridge, MA 02138, USA}

\author[0000-0002-7967-7676]{Rodrigo Córdova Rosado}
\affiliation{Center for Astrophysics $|$ Harvard \& Smithsonian,  60 Garden Street, Cambridge, MA 02138, USA}
\affiliation{Department of Astrophysical Sciences, Princeton University, 4 Ivy Lane, Princeton, NJ 08544, USA}

\author[0000-0002-9031-0904]{Sheperd S. Doeleman}
\affiliation{Center for Astrophysics $|$ Harvard \& Smithsonian,  60 Garden Street, Cambridge, MA 02138, USA}
\affiliation{Black Hole Initiative, Harvard University, 20 Garden Street, Cambridge, MA 02138, USA}

\author[0000-0003-4914-5625]{Joseph R. Farah}
\affiliation{Center for Astrophysics $|$ Harvard \& Smithsonian,  60 Garden Street, Cambridge, MA 02138, USA}
\affiliation{University of Massachusetts Boston, 100 William T, Morrissey Blvd, Boston, MA 02125, USA}

\author[0000-0002-4120-3029]{Michael D. Johnson}
\affiliation{Center for Astrophysics $|$ Harvard \& Smithsonian,  60 Garden Street, Cambridge, MA 02138, USA}
\affiliation{Black Hole Initiative, Harvard University, 20 Garden Street, Cambridge, MA 02138, USA}

\author[0000-0001-5461-3687]{Freek Roelofs}
\affiliation{Department of Astrophysics, Institute for Mathematics, Astrophysics and Particle Physics (IMAPP), Radboud University, P.O. Box 9010, 6500 GL Nijmegen, The Netherlands}

\author[0000-0002-6514-553X]{Remo P.J. Tilanus}
\affiliation{Department of Astronomy and Steward Observatory, University of Arizona, 933 N. Cherry Avenue, Tucson, AZ 85721, USA}

\author[0000-0002-4603-5204]{Jonathan Weintroub}
\affiliation{Center for Astrophysics $|$ Harvard \& Smithsonian,  60 Garden Street, Cambridge, MA 02138, USA}
\affiliation{Black Hole Initiative, Harvard University, 20 Garden Street, Cambridge, MA 02138, USA}

\correspondingauthor{Alexander W. Raymond}

\begin{abstract}
The Event Horizon Telescope (EHT) is a very long baseline interferometer built to image supermassive black holes on event-horizon scales. In this paper, we investigate candidate sites for an expanded EHT array with improved imaging capabilities.  We use historical meteorology and radiative transfer analysis to evaluate site performance.  Most of the existing sites in the EHT array have median zenith opacity less than 0.2 at 230~GHz during the March/April observing season.  Seven of the existing EHT sites have 345~GHz opacity less than 0.5 during observing months. Out of more than forty candidate new locations analyzed, approximately half have 230~GHz opacity comparable to the existing EHT sites, and at least seventeen of the candidate sites would be comparably good for 345~GHz observing.  A group of new sites with favorable transmittance and geographic placement leads to greatly enhanced imaging and science on horizon scales.
\end{abstract}

\section{Introduction}

The Event Horizon Telescope (EHT) is a very long-baseline interferometry (VLBI) array operating at 230~GHz~\citep{Doeleman2009,EHT2}.  In 2017, eight telescopes at six sites participated in the EHT observing campaign that yielded the first horizon-scale images of the supermassive black hole in the M87 galaxy (or \m87;~\citet{EHT1}\footnote{This black hole is also named \textit{P\={o}wehi} by A Hui He Inoa~\citep{kimura2019}}).  The planned 2021 EHT array includes 10 telescopes at nine sites plus the the Atacama Large Millimeter/submillimeter Array (ALMA): the Atacama Pathfinder Experiment (APEX), the Greenland Telescope (GLT), the Instituto de Radioastronom\'ia Milim\'etrica 30~m telescope (IRAM-30~m), the James Clerk Maxwell Telescope (JCMT), the 12~m radio telescope at Kitt Peak (KP) operated by the University of Arizona, the Large Millimeter Telescope Alfonso Serrano (LMT), the Northern Extended Millimeter Array (NOEMA), the Submillimeter Array (SMA), the Submillimeter Telescope (SMT), and the South Pole Telescope (SPT).  The coordinates for these stations are listed in Table~\ref{table:location_existing}.  Generally, these stations were developed as standalone submillimeter facilities and perform non-VLBI observations throughout most of the year.

There is an active effort~\citep{Blackburn2019a} to expand the array to new sites and higher frequency (e.g., 345~GHz), which will improve imaging and modeling of \m87 and the other primary science target, \sgra, the black hole candidate at the center of the Milky Way~\citep{Doeleman2008}.  These improvements will lead to a new instrument, the next-generation EHT (ngEHT), that is designed to deliver transformative science capabilities.  In this paper, we identify a set of candidate ngEHT sites, characterize their meteorological suitability for 230 and 345~GHz observations, and evaluate the additional VLBI baseline coverage they provide beyond the existing EHT array.  New analysis for the existing EHT sites is also presented.

There are many factors to consider in the selection of new sites for ngEHT stations.  The elevation of the primary science targets above the horizon, mutual visibility of those sources with the existing sites in the array, and the incremental Fourier coverage contributed by a new site are all key considerations.  New Fourier coverage improves the VLBI instrument in several ways.  Dense sampling improves imaging fidelity, and baseline redundancy in dense arrays is a powerful tool for calibration~\citep{Pearson_1984}.  A sufficiently dense array may even allow imaging to rely on calibration-independent closure quantities (three- and four-station products) instead of visibilities (two-station products; \citet{Chael2018b,Blackburn2019b}).  Finally, an array producing dense sampling is robust against losing one or two stations because of poor weather on a particular day.

The meteorological conditions are a key consideration in the evaluation of a new site.  Water is the atmospheric constituent that primarily governs atmospheric transmission and brightness temperature at millimeter and submillimeter wavelengths. We characterize precipitable water vapor (PWV, expressed in millimeter units) and liquid water path (LWP, expressed in microns) in our analysis.  Although submillimeter site evaluations usually emphasize the importance of PWV statistics, clouds consisting of liquid water droplets can also contribute significantly to opacity~\citep{Matsushita2003}.  Analysis of EHT data indicates that the imaging capability of the ngEHT will depend primarily on increasing the number of VLBI baselines, and only secondarily on sensitivity, so lower altitude sites with variable atmospheric conditions will be considered viable.  Our interest in sites that achieve good Fourier coverage motivates our analysis of meteorological conditions at suboptimal places, and distinguishes the present study from previous submillimeter site surveys, ~e.g.,~\citet{Tremblin2012}.

Historically, EHT observations have been scheduled for March or April.  Our analysis specifically addresses observing conditions during those months, which are the time of year when the sources are above the horizon at nighttime (the time of day when observing conditions are generally best for most sites), and when PWV paths are favorable across the Northern and Southern hemispheres.

\begin{deluxetable}{cccccccc}[t]
\tablewidth{0.9\columnwidth}
\tablecaption{Locations of Existing Sites (2021) in the Event Horizon Telescope Array~\citep{EHT2}.}
\tablehead{\colhead{\textbf{Site}} & \colhead{\textbf{Location}} & \colhead{\textbf{Lat.~(\degr)}} & \colhead{\textbf{Lon.~(\degr)}} & \colhead{\textbf{Alt.~(m)}} \\
 & \textbf{(Region, Country)} & & &} 
\startdata
ALMA & Antofagasta, CL & -23.03 & -67.75 & 5070 \\
APEX & Antofagasta, CL & -23.01 & -67.76 & 5100 \\
GLT\tablenotemark{a} & Avannaata, GL & 76.54 & -68.69 & 90 \\
IRAM-30m & Granada, ES & 37.07 & -3.39 & 2920 \\
JCMT & Hawaii, US & 19.82 & -155.48 & 4120 \\
KP & Arizona, US & 31.96 & -111.61 & 1900 \\
LMT & Puebla, MX & 18.98 & -97.31 & 4600 \\
NOEMA & Pr.-Alpes-C\^ote \\ & d'Azur, FR & 44.63 & 5.91 & 2620 \\
SMA & Hawaii, US & 19.82 & -155.48 & 4110 \\
SMT & Arizona, US & 32.70 & -109.89 & 3160 \\
SPT\tablenotemark{b} & South Pole, & -90.00 & 45.00 & 2820 \\
 & Antarctica & & & \\
\hline
\enddata
\tablenotetext{a}{site cannot observe \sgra}
\tablenotetext{b}{site cannot observe \m87}
\end{deluxetable}
\label{table:location_existing}

\begin{figure*}[ht]
\centering
  \includegraphics[width=0.8\linewidth]{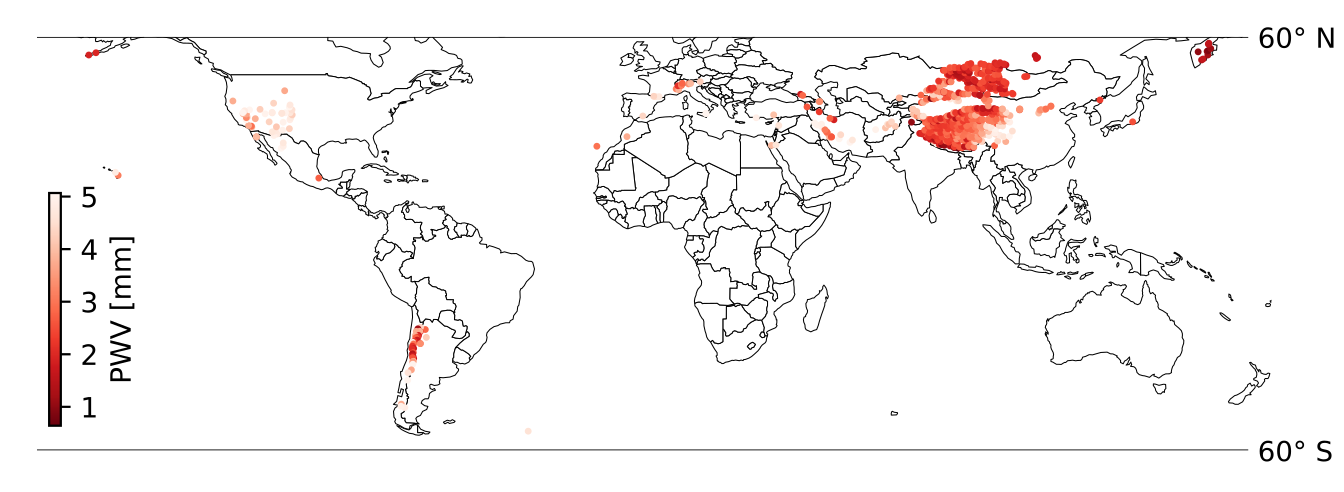}
  \caption{Worldwide March PWV for peaks above 2000~m in each 1\degr~latitude/longitude square between 60\degr~S and 60\degr~N latitude.  MERRA-2 datasets for years 2009 through 2012 were averaged.  A cutoff of 5~mm is applied.}
  \label{fig:PWVmap}
\end{figure*}

There are relatively few undeveloped locations for new telescopes in dry places that provide new VLBI baselines of hundreds to thousands of kilometers, which is the station spacing required to fill gaps in the existing EHT coverage.  To illustrate this point, we divide the globe into 1\degr~latitude by 1\degr~longitude squares, and within each square, the PWV and LWP values are calculated for the location of the highest peak based on the Modern-Era Retrospective analysis for Research and Applications, version 2 (MERRA-2;~\citet{Gelaro2017}). In Fig.~\ref{fig:PWVmap}, peaks are highlighted between 60$^\circ$ latitudes that have median PWV and LWP values less than 5~mm and 5~$\mu$m, respectively.  The locations of the dry peaks agree with previous PWV maps, e.g.,~\citet{Suen2015}, which have shown that the driest places tend to be nonequatorial high planes.  Maunakea, Parque Nacional Pico de Orizaba, the Atacama Desert, and other places hosting existing submillimeter telescopes emerge on the plot.  The Himalayas, Alps, Andes, and southwest United States each have many low-PWV sites.  The latter three regions have sustained covisibility of the primary science targets with ALMA, which is the most sensitive submillimeter station.  The large signal-to-noise ratio (S/N) achieved on ALMA baselines enables the atmospheric phase correction that improves and calibrates detections on non-ALMA baselines~\citep{EHT3}.  Selecting sites that have covisibility with the existing EHT stations also capitalizes on the $N\left(N-1\right)/2$ growth in the number of baselines, where $N$ is the number of stations that can simultaneously observe a given source.

\begin{figure*}[ht]
\centering
  \includegraphics[width=0.8\linewidth]{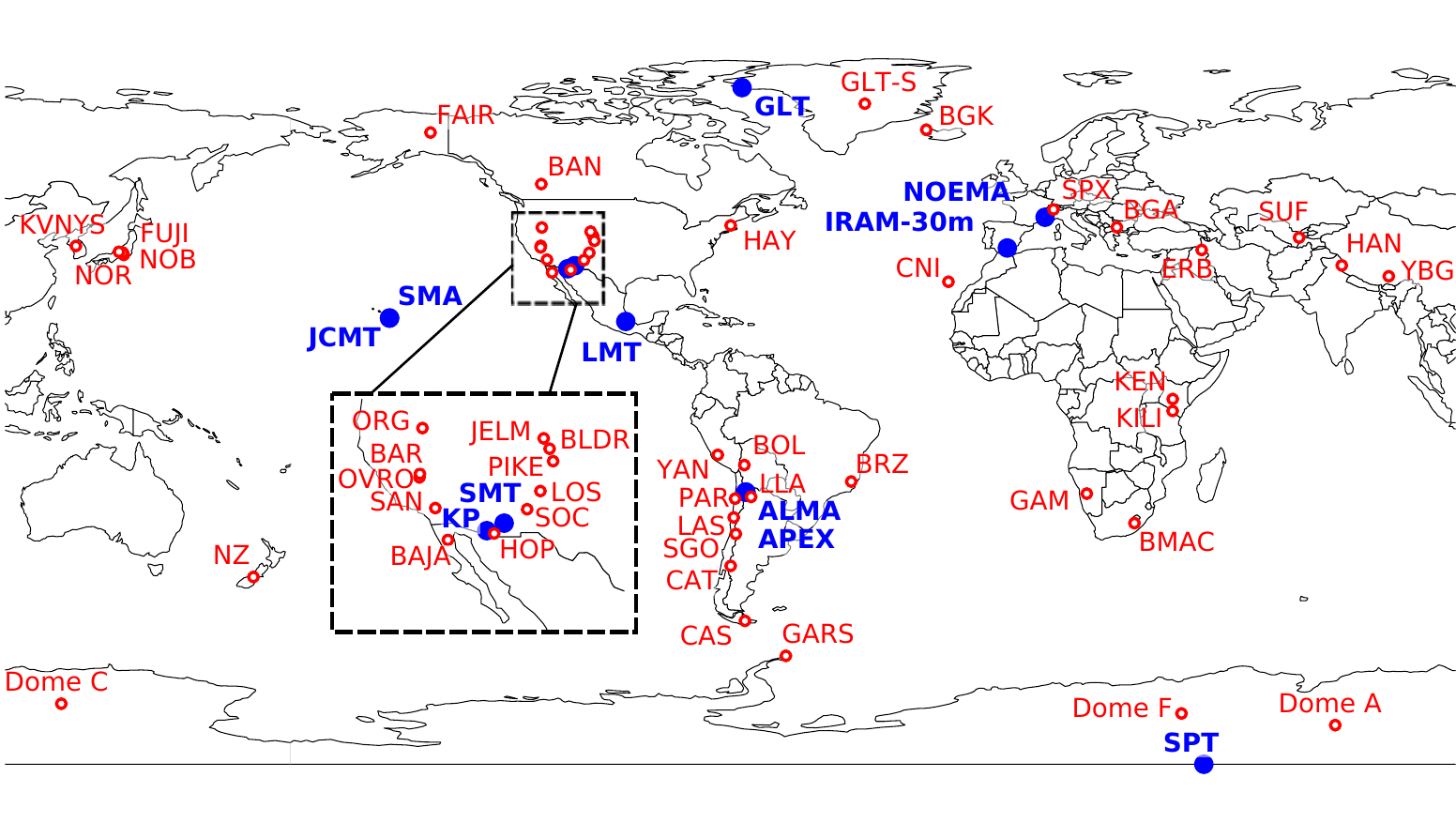}
  \caption{Map of planned 2021 EHT sites (blue, bold) listed in Table~\ref{table:location_existing} and 45 potential new sites (red) for the ngEHT listed in Table~\ref{table:location_new}.  The map is centered on the 2021 EHT sites.}
  \label{fig:map}
\end{figure*}

Using the PWV map as a guide together with knowledge about existing infrastructure and previous astronomical site evaluations, e.g., \citet{Suen2014}, we identify 45 sites to study in detail.  Those sites are plotted in Fig.~\ref{fig:map} and their longitude, latitude, and altitude coordinates are listed in Table~\ref{table:location_new}.  These sites cover a broad sampling of longitudes and latitudes, regions, and climates.  Our main objective in this paper is to catalog potential sites based primarily on weather and Fourier coverage.  The analysis we present is an initial step toward array expansion.  Many of the places we analyze already host an observatory or existing roads and infrastructure; however, some of the sites we present are remote and would require responsible assessments of the environmental and cultural impact of new construction to determine if development is viable~\footnote{The FUJI, KEN, and KILI sites we analyze in this study are UNESCO World Heritage Sites.  In each case the site has either been previously considered or used for astronomical purposes~\citep{Sekimoto2000,Graham2016}.  At the KILI site there are local efforts to explore the location for possible astronomical research (private communication, Dr. Noorali T. Jiwaji).  We present these sites to assess their meteorological viability but acknowledge that considerations beyond just the scientific suitability will determine the future possibilities for ngEHT telescopes at these locations.}.

We perform a radiative transfer analysis of the sites using historical meteorological statistics.  From that analysis, we present transmittances at 230~GHz and higher frequencies up to 500~GHz.  Observations at 345~GHz are of particular interest as a way to improve angular resolution and reduce scattering by the interstellar medium on the line of sight to \sgra. In general, 345~GHz observations demand sites with better weather than sites that are suitable for 230~GHz observations. Our calculations show which sites, new and existing, are feasible for high-frequency observing.

We evaluate the Fourier coverage contributed by the candidate stations in a few ways.  First, we present the geometric Fourier coverage contributed by each station while paying particular attention to the baselines with ALMA.  Second, we use the meteorological statistics for each site to calculate the probabilistic Fourier coverage from a set of Monte Carlo observing trials.  Finally, we perform test imaging reconstructions of \m87 and \sgra simulations to illustrate how new stations would enhance the imaging performance of the EHT.  We find that a collection of sites chosen for their favorable probabilistic Fourier coverage and good conditions is able to significantly improve the fidelity of the reconstructions.

\begin{deluxetable}{cccccccc}
\tablewidth{0.9\columnwidth}
\tablecaption{Next-generation Event Horizon Telescope Candidate Sites.}
\tablehead{\colhead{\textbf{Site}} & \colhead{\textbf{Location}} & \colhead{\textbf{Lat.~(\degr)}} & \colhead{\textbf{Lon.~(\degr)}} & \colhead{\textbf{Alt.~(m)}} \\
 & \textbf{(Region, Country)} & & &}
\startdata
BAJA & Baja California, MX & 31 & -115 & 2500 \\
BAN\tablenotemark{a} & Alberta, CA & 53 & -118 & 2000 \\
BAR & California, US & 38 & -118 & 3500 \\
BGA & Sofia, BG & 42 & 24 & 2500 \\
BGK\tablenotemark{a} & Westfjords, IS & 66 & -23 & 500 \\
BLDR & Colorado, US & 40 & -105 & 2500 \\
BMAC & Eastern Cape, ZA & -31 & 28 & 2500 \\
BOL & La Paz, BO & -16 & -68 & 5000 \\
BRZ & Esp\'irito Santo, BR & -20 & -42 & 2500 \\
CAS & Tierra del Fuego, AR & -55 & -68 & 500 \\
CAT & R\'io Negro, AR & -41 & -71 & 2000 \\
CNI & La Palma, ES-CN & 29 & -18 & 2000 \\
Dome A\tablenotemark{b} & Antarctica & -80 & 77 & 4000 \\
Dome C\tablenotemark{b} & Antarctica & -75 & 123 & 3000 \\
Dome F\tablenotemark{b} & Antarctica & -78 & 39 & 3500 \\
ERB & Kurdistan, IQ & 37 & 44 & 2000 \\
FAIR\tablenotemark{a} & Alaska, US & 65 & -145 & 1000 \\
FUJI & Fujinomiya \& \\ & Yamanashi, JP & 35 & 139 & 3500 \\
GAM & Khomas, NA & -23 & 16 & 2000 \\
GARS & Antarctica & -63 & -58 & 0 \\
GLT-S\tablenotemark{a} & Northeastern, GL & 73 & -38 & 3000 \\
HAN & Jammu \& Kashmir, IN & 33 & 79 & 4000 \\
HAY & Massachusetts, US & 43 & -71 & 0 \\
HOP & Arizona, US & 32 & -111 & 2000 \\
JELM & Wyoming, US & 41 & -106 & 2500 \\
KEN & Meru, KE & -0 & 37 & 4000 \\
KILI & Kilimanjaro, TZ & -3 & 37 & 4500 \\
KVNYS & Seoul, KR & 38 & 127 & 0 \\
LAS & Coquimbo, CL & -29 & -71 & 2000 \\
LLA & Salta, AR & -24 & -66 & 4500 \\
LOS & New Mexico, US & 36 & -106 & 2000 \\
NOB & Nagano, JP & 36 & 138 & 1000 \\
NOR & Gifu, JP & 36 & 138 & 2500 \\
NZ & Canterbury, NZ & -44 & 171 & 2000 \\
ORG & Oregon, US & 42 & -118 & 2000 \\
OVRO & California, US & 37 & -118 & 1000 \\
PAR & Antofagasta, CL & -25 & -70 & 2500 \\
PIKE & Colorado, US & 39 & -105 & 4000 \\
SAN & California, US & 34 & -117 & 2500 \\
SGO & Santiago, CL & -33 & -70 & 3500 \\
SOC & New Mexico, US & 34 & -108 & 2000 \\
SPX & Bern, CH & 47 & 8 & 3500 \\
SUF & Jizzakh, UZ & 40 & 68 & 2000 \\
YAN & Huanca Sancos, PE & -14 & -75 & 4500 \\
YBG & Yangbajing Tibet, CN & 30 & 91 & 4000 \\
\enddata
\tablenotetext{a}{site cannot observe \sgra}
\tablenotetext{b}{site cannot observe \m87}
\end{deluxetable}
\label{table:location_new}

\section{Methods}
We use atmospheric state data from MERRA-2 as inputs to our calculations.  Each site is analyzed using a 10 year historical dataset ending on 2019 January 1.  The datasets consist of 3~hr averages, gridded to 0.5\degr~latitude and 0.625\degr~longitude and 42 vertical pressure levels.  Each time, position, and altitude datum contains the state of the atmosphere including water quantities (vapor, liquid, and ice) and temperature.  The gridded data are interpolated to the precise location of a candidate site.  We calculate water statistics as well as submillimeter radiative properties with the aid of the \textit{am} atmospheric modeling and radiative transfer code~\citep{Paine2019}.  For each site, we compute 10 year monthly statistics of column-integrated water vapor and liquid clouds, and we use those to calculate the spectral opacity $\tau \left( \nu \right)$ and brightness temperature $T_\mathrm{b}$ as a function of frequency $\nu$ following the methodology described in \citet{Paine2018,Paine2019}.  The liquid and ice opacity calculations utilize the Rayleigh approximation.

\begin{figure}[t]
\centering
  \includegraphics[width=\columnwidth]{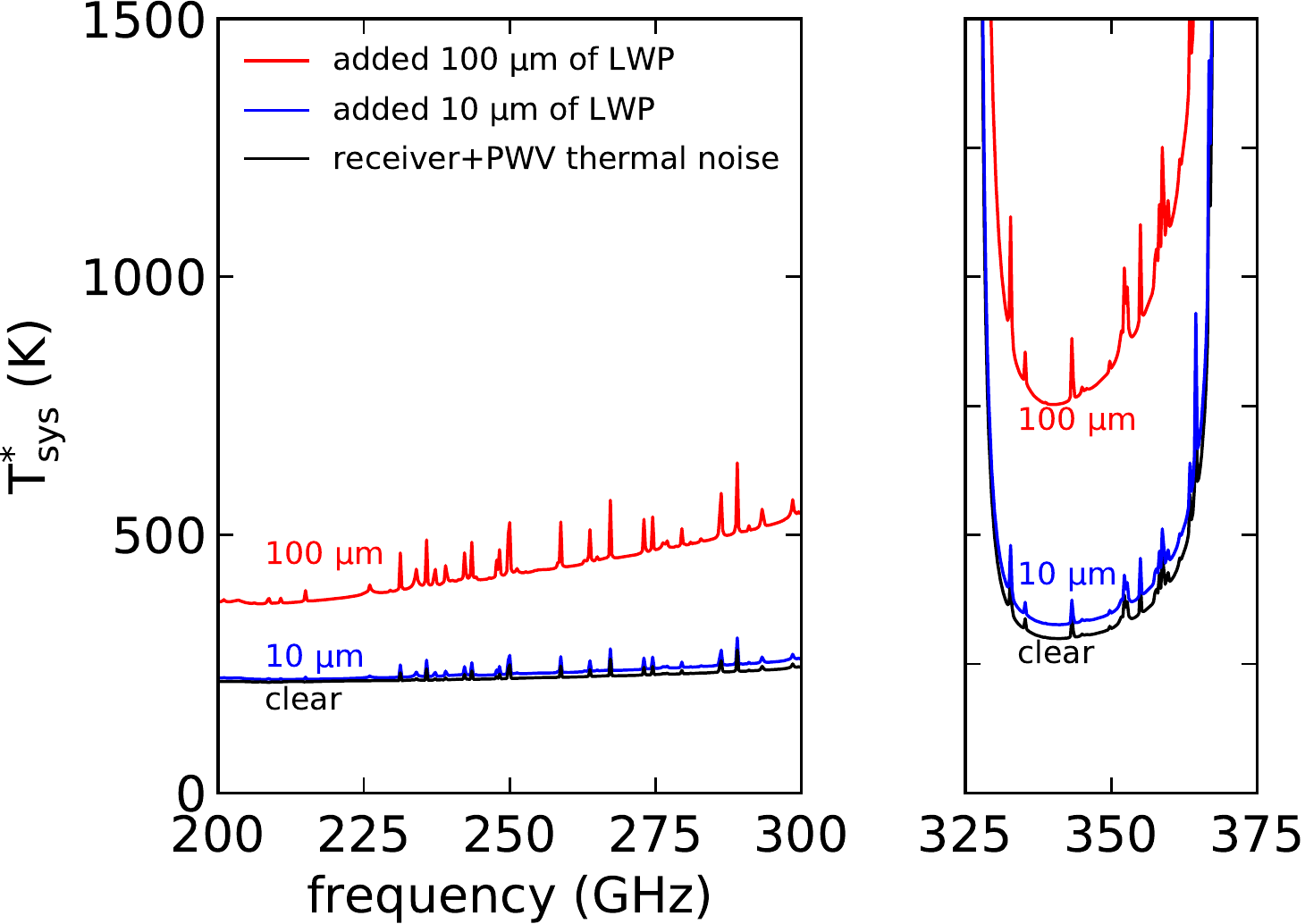}
  \caption{Effective system temperature along a line of sight at 20$\degr$ elevation for the ALMA/APEX site. The indicated liquid water path is added to the meteorological conditions of the April clear-sky days (i.e., to the days with less than 10~$\mu$m LWP) at the 450~mbar pressure level.  The liquid component can double the effective system temperatures at 230 or 345~GHz.}
  \label{fig:LWP}
\end{figure}

Submillimeter wavelength observations are affected by meteorology in several ways, and we consider various factors in our site evaluation.  First, PWV statistics are frequently used as proxies for submillimeter observing conditions, and we present monthly PWV statistics for both existing EHT and candidate ngEHT sites.  The LWP statistic is another important metric we consider.  The LWP contribution to zenith opacity will be approximately $2.5\times10^{-3}$ and $3.5\times10^{-3}$\,$\mu$m\textsuperscript{--1} at 230 and 345~GHz, respectively, with weak temperature dependence~\citep{Paine2019}.  In other words, 100~$\mu$m of liquid water path contributes 0.25 to the opacity at 230~GHz.  Figure~\ref{fig:LWP} shows the 230~GHz effective system temperature (defined later in this section) at a typical lower bound on telescope elevation angle during VLBI observations; at 100~$\mu$m LWP column, liquid clouds become the dominant source of system noise.  Consequently, we include the effects of all water phases in our radiative transfer analysis.

An additional metric that is important in interferometry is the coherence time of the atmosphere.  Turbulence cells in the moist atmosphere cause phase fluctuations in the interferometric response of a VLBI baseline resulting in signal loss for extended coherent integrations.  Here, we define the coherence time to be the integration interval beyond which the VLBI signal loss rises above 10\%.  During the 2017 EHT observations, the coherence time was typically in the range of 10--20~seconds on ALMA baselines~\citep{EHT2}.  Strong S/N detections to ALMA are used to correct the phase data and enable longer coherent integration time. The spatiotemporal grid for MERRA-2 is not suitably fine for calculating coherence times, so we do not consider that effect in this analysis; however, atmospheres with small PWV and LWP characteristics will generally have small wet path delays and consequently better coherence times~\citep{TMS2017}.

We calculate the incremental Fourier coverage, or Fourier filling, of the \m87 and \sgra sources for new sites compared to a fiducial array, i.e., the 2021 EHT array.  As the Earth rotates during an observation, each baseline samples a set of Fourier components that traces tracks in the \textit{u}-\textit{v} (east--west and north--south) Fourier plane.  The incremental Fourier coverage is the fractional increase in the \textit{u}-\textit{v} area sampled by an additional station compared to the fiducial array. The width of each each track is given by $0.71$ divided by the field of view (FoV) adopted for imaging, which corresponds to the half-width of the visibility response to a filled disk occupying the FoV on the sky~\citep{Palumbo2019}.  If a new site primarily samples Fourier space that is already covered by existing baselines, then its incremental coverage is small.  If a new site samples spatial frequencies that are not covered by the existing array, then its incremental coverage will be large.  There are plans to position an antenna at the Greenland Summit, which has excellent atmospheric conditions~\citep{Matsushita2017}.  We assume that the GLT site remains in the fiducial array and evaluate the incremental addition of a GLT-S station.

To incorporate the effects of weather in the Fourier filling metric, we perform 45 repeated Monte Carlo trial simulations of the weather-dependent observations using \texttt{eht-imaging}~\citep{Chael2016,Chael2018b} based on the statistical distribution of opacity.  In each trial, we flag low S/N portions of the Fourier coverage (S/N~<~3, in accordance with the long baseline scans reported in \citet{EHT3}), and report the average coverage across all trials.  We make some simplifying assumptions like neglecting losses resulting from the beam pointing off source or mechanical defocusing.  Those effects could be incorporated in the future using synthetic data tools developed for the EHT~\citep{Roelofs2020}.  The \m87 source model we use in the simulations has 0.6~Jy flux density at 230~GHz on the spatial scale of the event horizon.  A scattering kernel is applied to the \sgra source model to simulate the effects of the interstellar medium \citep{Johnson_2018}, and a short baseline 230~GHz flux density of 2.3~Jy is used for that source.

The simulated S/N values for each synthetic observation are governed by the system-equivalent flux density for each site ($\mathrm{SEFD}_i$),
\begin{equation}
\mathrm{SEFD}_i =2k T_{\mathrm{sys},i}^\ast/A_{\mathrm{eff},i},
\end{equation}
where $A_{\mathrm{eff},i}$ is the effective antenna area and $k$ is the Boltzmann constant.  For the 2021 array as well as the new sites HAY, KVNYS, LAS, NOB, SUF, and GAM, we base the collecting area on the antenna diameter of the existing or planned telescopes, which are 37~\citep{Rogers1993,Usoff2014}, 21~\citep{Lee2011}, 15~\citep{booth1989}, 45~\citep{Ishiguro1986}, 70, and 15~m, respectively.  For all other sites, we assume a 10~m diameter antenna. Known aperture efficiencies are used, and for the new 10~m stations we assume a Ruze-formula~\citep{TMS2017} aperture efficiency based on a 64~$\mu$m rms surface accuracy, which is the quadrature sum of a 40~$\mu$m primary surface accuracy and an effective 50~$\mu$m focus offset.  The $\mathrm{SEFD}_i$ values depend on the effective system temperature, $T_{\mathrm{sys},i}^\ast = \left[ T_\mathrm{R} + T_{\mathrm{b},i} \left( \theta \right) \right] e^{\tau_i \left(\theta \right)}$~\citep{TMS2017}, and the opacity for a given site.  $\tau_i \left(\theta \right)$ is the opacity toward the source as a function of elevation, $\theta$, where a lower limit of 10$^{\circ}$ in elevation is applied.

We adopt a receiver temperature, $T_\mathrm{R}$, of 60 and 100~K for all sites at 230 and 345~GHz, respectively, and use the interquartile ranges for sky-brightness temperature ($T_{\mathrm{b},i}$) and opacity ($\tau_i$) to generate a Gaussian random variable for $T_{\mathrm{sys},i}^\ast$. The random variables at the ALMA/APEX, JCMT/SMA, and KP/SMT EHT sites are assumed to be correlated while those for candidate ngEHT sites are assumed to be uncorrelated.  The S/N of the visibility measured on a single baseline is~\citep{TMS2017}

\begin{equation}
    S/N = \eta \mathcal{S_{\nu}} \left(\vec{u} \right) \sqrt{\frac{2 \Delta \nu \tau_\mathrm{c}}{\mathrm{SEFD}_1 \mathrm{SEFD}_2 }},
\end{equation}

\noindent where $\eta$ is the digital and processing efficiency, $\mathcal{S_{\nu}} \left(\vec{u} \right)$ is the correlated flux density on baseline $\vec{u}$, $\Delta \nu$ is the bandwidth, $\tau_\mathrm{c}$ is the integration time.  We specify integration times of 300~s (\m87) and 100~s (\sgra, shorter than the smallest expected period of the innermost stable circular orbit). The bandwidth specified for S/N calculations is 2 (as for the current EHT) or 8~GHz (ngEHT), and corresponds to the Nyquist band of a single digitizer~\citep{EHT2}.  We call this the fringe-finding bandwidth.  Multiple fringe-finding sidebands are aggregated for imaging: two of 2~GHz each in 2017, four of 2~GHz each in 2021, and two of 8~GHz each for the ngEHT.  Those correspond to a total bandwidth across two polarizations of 8, 16, and 64~GHz in the respective years (assuming simultaneous dual-frequency 230+345~GHz observing with the ngEHT).  The dual-frequency approach increases the Fourier coverage, and with the appropriate model, the two frequencies can be combined to synthesize a single image~\citep{Sault1994}.   We account for the slightly different center frequencies of each sideband when calculating the Fourier filling.

Finally, we perform imaging reconstructions of \m87 and \sgra using a possible ngEHT array chosen by picking from among the best sites from the Fourier filling metric.  The reconstruction is meant as an example of how new sites will affect the imaging capabilities of the EHT instrument.  The source models for those reconstructions are general-relativistic magnetohydrodynamic (GRMHD) simulations of accretion flows~\citep{Rowan2017,Chael2018a,Chael2019}.  We calculate the noise for each baseline in the same manner described above under median opacity conditions.  For \m87, we generate synthetic data with a $40\%$ duty cycle, and for \sgra, we use a brief 100~s snapshot observation that begins at a time of day when the Fourier filing is greatest, which can be different for different arrays.  We apply an imaging process similar to the published \texttt{eht-imaging} script released alongside \citet{PaperIV}.

\section{Results and Discussion}
\subsection{Weather and Observing Statistics}
\begin{figure}
  \includegraphics[width=\columnwidth]{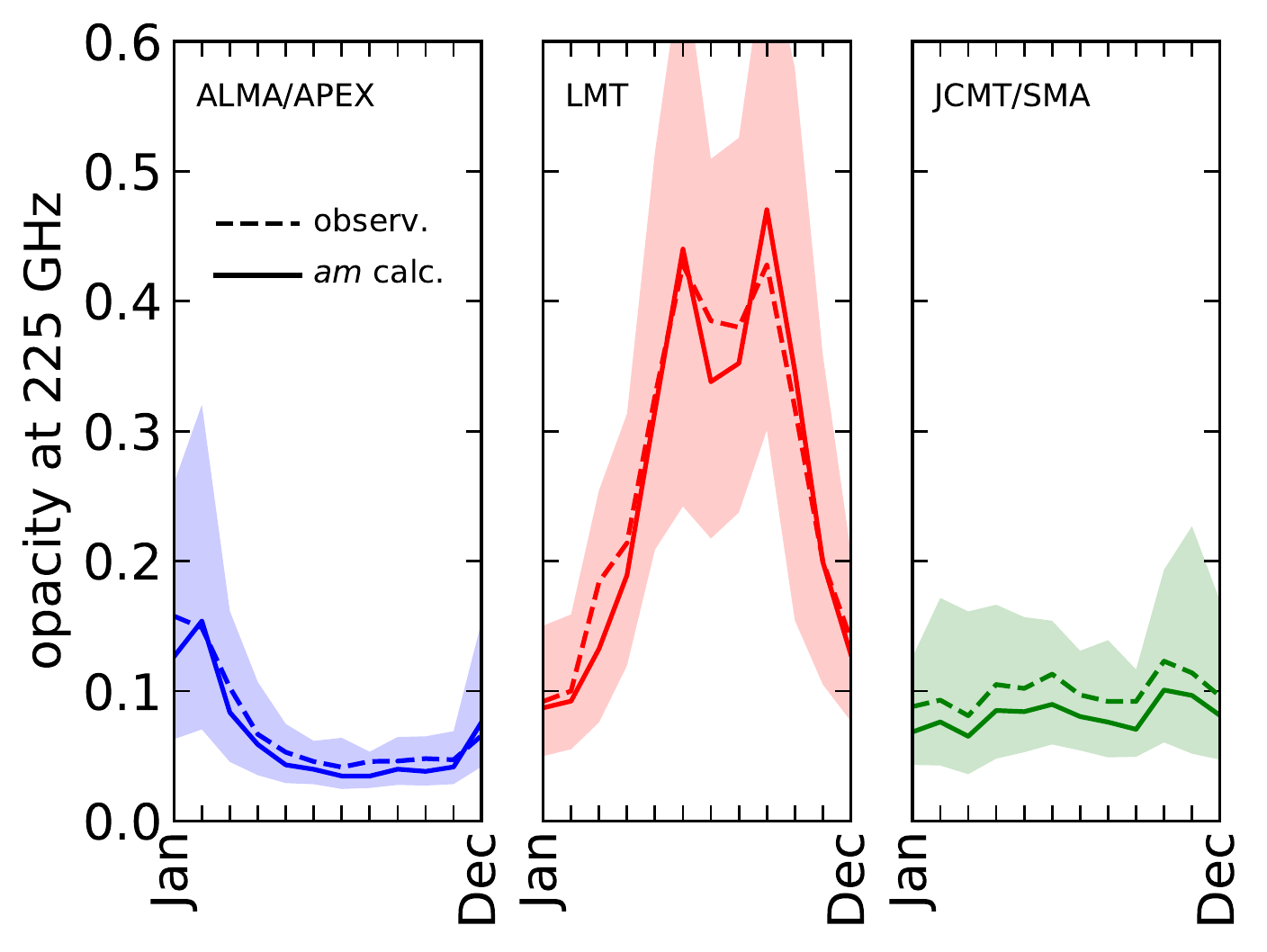}
  \caption{Monthly 225~GHz zenith opacity ($\tau_{225}$) for three existing EHT sites.  The calculated values (solid) from the reanalysis and 225~GHz radiometer observations (dashed) are comparable.  The shaded region bounds the interquartile range, i.e., the middle 50\% between the 25th (lower) and 75th (upper) percentiles for the reanalysis.}
  \label{fig:radiometer}
\end{figure}

Figure~\ref{fig:radiometer} compares the 225~GHz opacity calculated with the \textit{am} code using MERRA-2 data against the logged tipping radiometer measurements made at three different EHT sites.  The tipping radiometer datasets are each multiyear: 1995-2004 for ALMA/APEX\footnote{http://legacy.nrao.edu/alma/site/}, 2013-2017 for LMT~\citep{Ferrusca2014}, and 2009--2014 for SMA~\citep{Radford2016}.  The tipper data were minimally flagged for $\tau_{\mathrm{225}}$ extremes, and more than 95\% of entries were kept for each site.  The \textit{am} calculation is for the same dates as the tipper entries except for APEX, which is done for our nominal 10 year period ending on the first day of year 2019.  The agreement between the medians of the calculations and field measurements supports our methods.  The calculation reproduces the wide range of seasonal opacities between median $\tau_{\mathrm{225}}=0.04$ at APEX in July to $\tau_{\mathrm{225}}=0.47$ at LMT.  Seasonal features, like the dip in July/August opacity at the LMT, are also reproduced.  Figure~\ref{fig:radiometer} confirms that the calculation methods accurately reflect ensemble observing conditions at geographically varied sites.

\begin{figure*}[ht]
\centering
  \includegraphics[width=\linewidth]{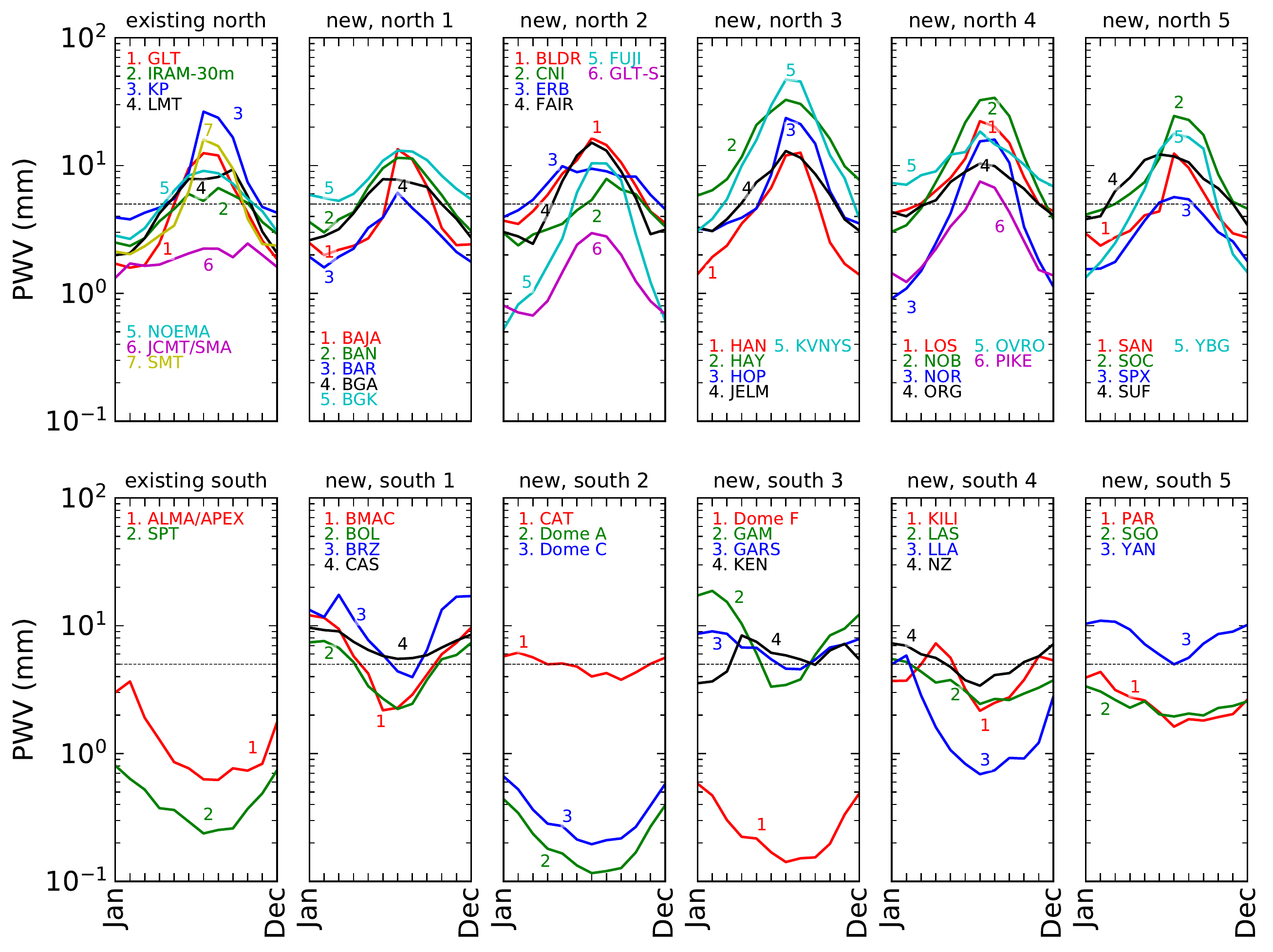}
  \caption{Median monthly precipitable water vapor during the year at existing and candidate new EHT sites.  Northern sites are in the top row and southern sites are in the bottom row.  In April, existing sites are below 5~mm, so a dotted line is added at that level.}
  \label{fig:PWVall}
\end{figure*}

Figure~\ref{fig:PWVall} plots the PWV statistics throughout the year at existing and candidate new sites.  The northern sites (top row) generally exhibit minimum PWV values during December and January.  Southern sites in the bottom row exhibit minimum PWV values during July and August.  Historically, EHT observations are performed in March or April, when the northern sites still have reasonably dry atmospheres and ALMA has a median PWV column of 1--2~mm.  All of the existing EHT sites have a median PWV column in March and April that is less than 5~mm.  Most of the sites we analyze have comparable PWV statistics in those months.

The PWV averages we report, which are derived from the MERRA-2 data, agree with measurements available in the literature for both the 2021 array sites (plus ALMA) and for the candidate new sites.  At the South Pole, which is presently the driest site in the array, radiosonde measurements collected over several decades show PWV ranges from about 0.25~mm in July to just below 0.8~mm in January~\citep{Chamberlin2001}.  The medians we derive for the SPT site are 0.24~mm in July and 0.8~mm in January.  At Kitt Peak, multiyear infrared measurements done in the 1980s found median PWVs ranging from about 4~mm in winter to 27~mm in August during monsoon season~\citep{Wallace1984}.  We find 3.8--4.2~mm in winter and about 23~mm in August.  Thus, for the driest and wettest sites in the EHT array, we find that the interpolated MERRA-2-based PWV agrees with what has been measured in the field.

PWV field measurements have also been made at many of the candidate new sites, and in most of those cases, we also find that the MERRA-2 PWV values are reliable. At Dome~A and Dome~C, the yearly median PWV values reported from satellite data from 2008 to 2010 are 0.21 and 0.28~mm, respectively~\citep{Tremblin2012}.  We find yearly medians of 0.19 and 0.29~mm for those sites, which are in good agreement as expected since the satellite data used would have been assimilated into MERRA-2.  Ground-based radiometer measurements between 2009 and 2014~\citep{Ricaud2015} observed mean seasonal PWV values of about 0.3~mm during June--August and about 0.7~mm in December--February.  We find a similar range: 0.2~mm in June--August and 0.6~mm in December--February.

Precipitable water vapor measurements for Teide and Roque de los Muchachos Observatory in the Canary Islands were made using Global Positioning System path delays and radiosondes~\citep{Castro-Almazan2016}.  For the years 2012 and 2013, monthly median PWV values at CI ranged from about 2~mm in February to about 9~mm in August compared with our 10 year medians of 2.3~mm and 7.9~mm in February and August, respectively.

The European Southern Observatory performed a multi-instrument campaign involving optical, far infrared, and radiosonde measurements at Paranal and La Silla~\citep{Querel2010}.  The study concluded that between 2005 and 2009, those sites had mean PWV columns of $2.3\pm1.8$ and $3.4\pm2.4$~mm, respectively.  Those values are close to the mean of the monthly medians we obtain from interpolating MERRA-2 data: 2.6~mm at PAR and 3.6~mm at LAS.  Both the \cite{Querel2010} result and our value agree with a two-year field measurement made using a 183~GHz radiometer~\citep{Kerber2015}, which found 3.0~mm.

While the agreement with the literature is good for most of the sites we analyze, there are a few cases where the agreement is not as close.  Far-infrared radiometer measurements made on intervals between 1984 and 1987 at Pico Veleta~\citep{Quesada1989} returned median PWV values of 1.2~mm in January up to 4.1~mm in September. We find slightly more water for IRAM-30~m: 2.5~mm in January and 6.7~mm in September.  At JELM, field measurements beginning in the late 1970s found 1.4~mm of water vapor in winter up to 6.8~mm in August~\citep{Grasdalen1985}.  We find about 3.0~mm in winter and almost 12~mm in August.  In these cases, discrepancies could be caused by local effects that the MERRA-2 profiles do not resolve.  Future field measurements will be needed to determine if such anomalies exist at particular sites.

\begin{figure}[ht]
\centering
  \includegraphics[width=\columnwidth]{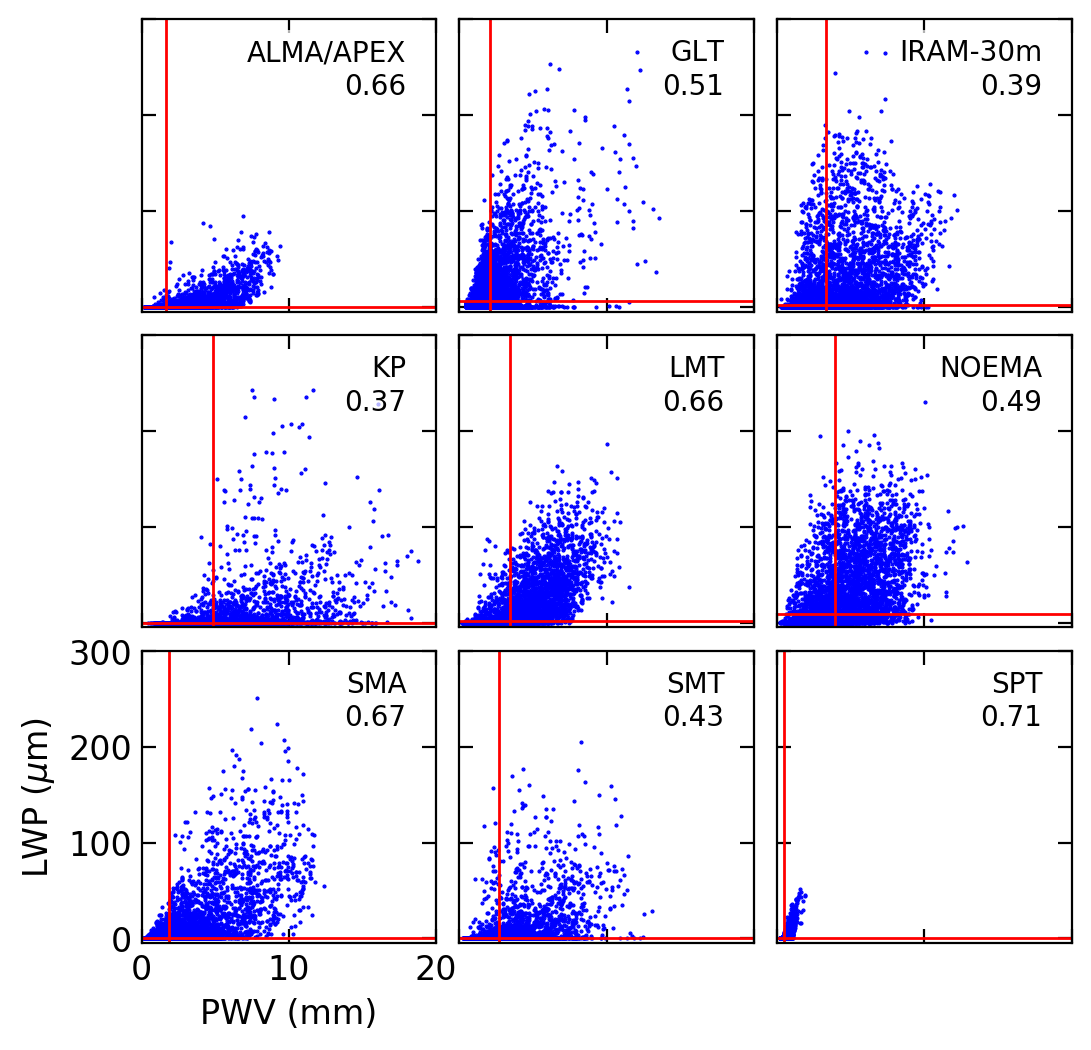}
  \caption{Precipitable water vapor and liquid water path for existing EHT sites during March and April.  The Pearson correlation coefficient between PWV and LWP is listed below the site label.  The vertical and horizontal lines show the median PWV and LWP, respectively.}
  \label{fig:scatterOLD}
\end{figure}

\begin{figure*}[ht]
\centering
  \includegraphics[width=1.5\columnwidth]{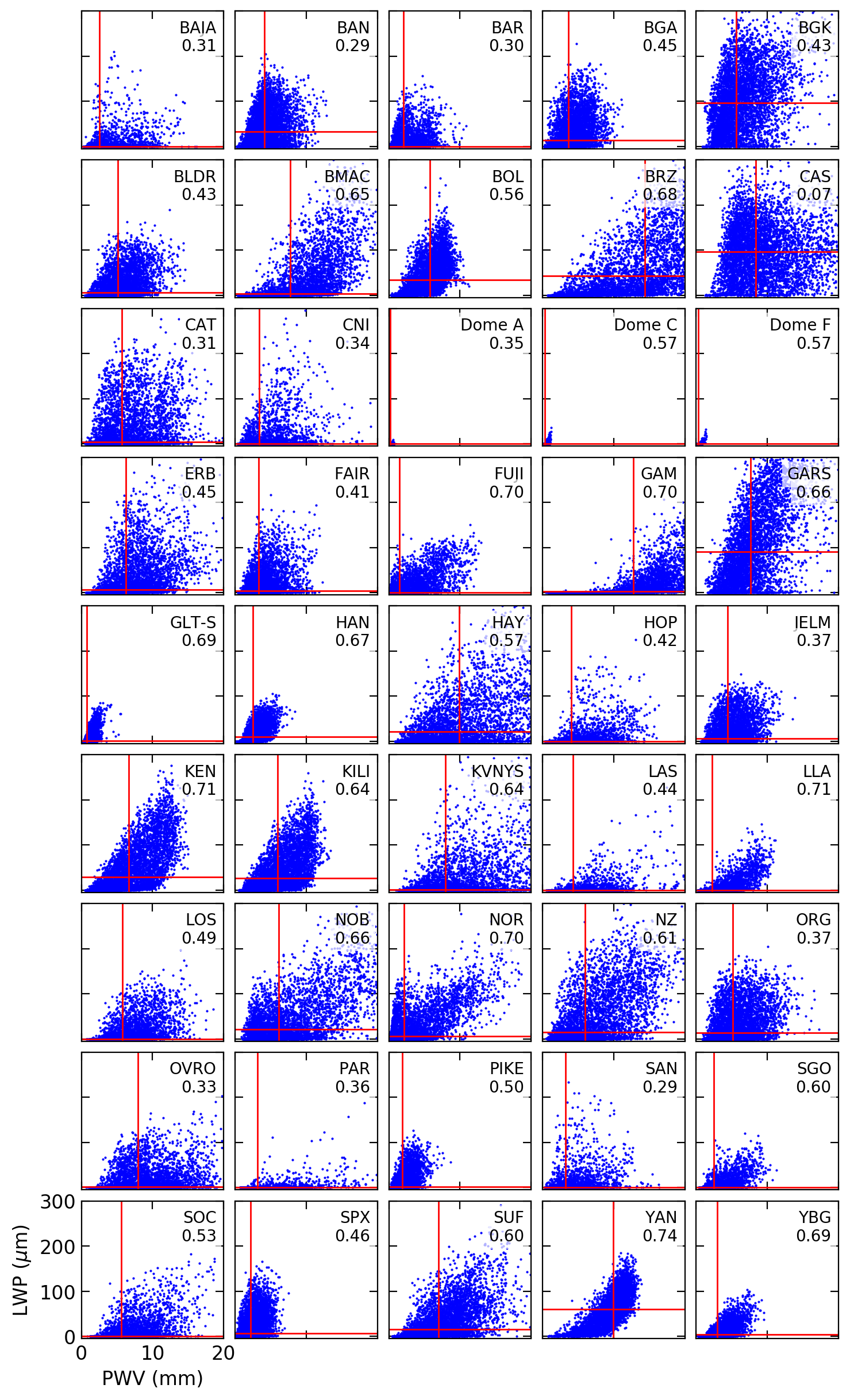}
  \caption{Same as Fig.~\ref{fig:scatterOLD} but for candidate new sites.}
  \label{fig:scatterNEW}
\end{figure*}

As we have already established, cloud liquid water also affects submillimeter opacity.  In Fig.~\ref{fig:scatterOLD}, the LWP and PWV values at the existing EHT sites are plotted for each 3~hr bin in the MERRA-2 dataset, and the median of each axis is marked with a red line.  The correlation between the LWP and PWV variables ranges from weak to strong depending on the site, with the Pearson coefficient varying from about 0.3 to a little more than 0.7, and the maximum LWP values range from 50~$\mu$m at the South Pole to approximately 200~$\mu$m at several of the sites. Figure~\ref{fig:scatterNEW} shows the same information but for the candidate new sites.  The LWP/PWV correlation coefficients have a similar range as the existing sites: approximately 0.3 to 0.75 with the exception of CAS.  At the BAN, BGA, BGK, NOR, NZ, and SPX sites, there are a significant number of days when the vapor content is less than 5~mm, but the liquid content is approaching 100~$\mu$m.

At some of the sites in Figs.~\ref{fig:scatterOLD} and \ref{fig:scatterNEW}, there is only moderate correlation between the LWP and PWV statistics.  The frequency of appreciable cloud cover is therefore important in its own right.  In Table~\ref{table:clouds}, we show the number of days per month with at least 50~$\mu$m of LWP column on a bimonthly basis, which corresponds to a 0.13 and 0.18 opacity increase above the clear-sky conditions at 230 and 345~GHz, respectively.  In March, the existing EHT sites have an average of five or fewer days each month with heavy cloud cover.  Twenty-five of the candidate new sites have a comparable frequency of clouded days and would therefore be similarly reliable for observing.

Unlike PWV statistics, which have a seasonality that depends strongly on hemisphere, the seasonality of cloud cover is more regional.  For example, the sites in the southwestern United States have increased cloudiness in late summer because of the North American monsoon.  Except for NOEMA, which does not exhibit much seasonal variation to begin with, none of the existing EHT sites has its greatest number of clouded days in March.  This supports the idea that March and April are good months for EHT observations.  Of the candidate new sites, some, like BGK and YAN, probably have a prohibitive number of clouded days throughout the year. Other sites show remarkable seasonal dependence.  For example, GAM goes from eight clouded days in January to zero days throughout June and August.  If the EHT eventually moves from once-per-year, campaign-style observations to remotely controlled observations that target the optimal conditions for each source, then it may be possible to capitalize on such variability.

While PWV and LWP are good indicators of observing conditions, radiative transfer at a particular site also depends on local vertical properties of moisture and temperature \citep{Paine2019} and must therefore be calculated.  Table~\ref{table:tau} presents the zenith 230~GHz opacity.  Opacities are calculated using the local thermodynamic conditions above each site.  Neither March nor April are extreme seasons for the 230~GHz zenith opacity.  Of the candidate new sites, 21 of them have median zenith opacity of less than 0.2 during either of those months.  The sites with median opacity below 0.10 (Dome~A, Dome~C, Dome~F, FUJI, GLT-S, LLA, NOR, and PIKE) have interquartile ranges that vary from less than 0.01 at Antarctic sites to 0.2 elsewhere.  A second tier of sites have opacities between 0.1 and 0.15 (BAJA, BAR, HAN, PAR, SGO, SPX, and YBG).

\begin{deluxetable}{ccgcccc}
\tabletypesize{\footnotesize}
\tablecaption{Days with at Least 50~$\mu$m Liquid Water Path.}
\tablehead{\colhead{\textbf{Site}}  & \colhead{\textbf{Jan.}} & \colhead{\textbf{Mar.}} & \colhead{\textbf{May}}& \colhead{\textbf{Jul.}} & \colhead{\textbf{Sep.}}& \colhead{\textbf{Nov.}}} 
\startdata
ALMA/APEX & $1^{2}_{0}$ & $0^{1}_{0}$ & $0^{0}_{0}$ & $0^{0}_{0}$ & $0^{0}_{0}$ & $0^{0}_{0}$ \\
GLT & $2^{4}_{1}$ & $2^{4}_{1}$ & $9^{11}_{8}$ & $10^{13}_{9}$ & $13^{16}_{12}$ & $7^{9}_{6}$ \\
IRAM-30m & $2^{4}_{2}$ & $4^{6}_{3}$ & $3^{5}_{2}$ & $0^{0}_{0}$ & $2^{4}_{2}$ & $4^{6}_{3}$ \\
JCMT/SMA & $1^{2}_{1}$ & $3^{5}_{2}$ & $1^{3}_{1}$ & $0^{1}_{0}$ & $2^{3}_{1}$ & $2^{4}_{2}$ \\
KP & $2^{4}_{2}$ & $1^{2}_{0}$ & $0^{1}_{0}$ & $9^{11}_{8}$ & $4^{6}_{3}$ & $1^{2}_{1}$ \\
LMT & $0^{1}_{0}$ & $2^{3}_{1}$ & $7^{10}_{6}$ & $10^{12}_{9}$ & $15^{18}_{14}$ & $2^{4}_{2}$ \\
NOEMA & $4^{5}_{3}$ & $5^{7}_{4}$ & $10^{12}_{9}$ & $6^{8}_{5}$ & $6^{8}_{5}$ & $5^{7}_{4}$ \\
SMT & $2^{3}_{1}$ & $1^{2}_{1}$ & $1^{2}_{0}$ & $10^{12}_{9}$ & $6^{8}_{5}$ & $1^{2}_{1}$ \\
SPT & $0^{0}_{0}$ & $0^{0}_{0}$ & $0^{0}_{0}$ & $0^{0}_{0}$ & $0^{0}_{0}$ & $0^{0}_{0}$ \\
\\
\hline
BAJA & $1^{3}_{1}$ & $0^{1}_{0}$ & $0^{1}_{0}$ & $3^{4}_{2}$ & $2^{4}_{2}$ & $0^{1}_{0}$ \\
BAN & $6^{8}_{5}$ & $11^{13}_{9}$ & $13^{15}_{11}$ & $16^{19}_{15}$ & $11^{13}_{9}$ & $9^{11}_{7}$ \\
BAR & $1^{2}_{0}$ & $1^{2}_{1}$ & $3^{5}_{2}$ & $3^{5}_{3}$ & $1^{2}_{1}$ & $0^{1}_{0}$ \\
BGA & $3^{4}_{2}$ & $6^{8}_{5}$ & $13^{15}_{12}$ & $10^{12}_{9}$ & $7^{9}_{6}$ & $3^{5}_{2}$ \\
BGK & $23^{25}_{22}$ & $23^{25}_{21}$ & $19^{22}_{18}$ & $16^{18}_{15}$ & $22^{24}_{21}$ & $22^{24}_{21}$ \\
BLDR & $2^{4}_{1}$ & $3^{5}_{3}$ & $7^{9}_{6}$ & $7^{9}_{6}$ & $5^{6}_{4}$ & $2^{3}_{1}$ \\
BMAC & $10^{13}_{9}$ & $8^{10}_{6}$ & $2^{4}_{1}$ & $1^{2}_{0}$ & $1^{2}_{1}$ & $4^{6}_{3}$ \\
BOL & $17^{19}_{15}$ & $13^{16}_{12}$ & $5^{7}_{4}$ & $3^{4}_{2}$ & $8^{10}_{7}$ & $10^{12}_{9}$ \\
BRZ & $13^{16}_{12}$ & $17^{19}_{15}$ & $7^{9}_{6}$ & $3^{5}_{3}$ & $5^{7}_{4}$ & $16^{18}_{15}$ \\
CAS & $26^{28}_{26}$ & $23^{26}_{22}$ & $19^{21}_{17}$ & $17^{19}_{16}$ & $18^{21}_{17}$ & $24^{26}_{23}$ \\
CAT & $5^{7}_{4}$ & $6^{8}_{5}$ & $9^{12}_{8}$ & $8^{10}_{7}$ & $8^{10}_{7}$ & $8^{10}_{7}$ \\
CNI & $1^{2}_{1}$ & $2^{3}_{1}$ & $0^{0}_{0}$ & $0^{0}_{0}$ & $0^{1}_{0}$ & $2^{4}_{2}$ \\
Dome A & $0^{0}_{0}$ & $0^{0}_{0}$ & $0^{0}_{0}$ & $0^{0}_{0}$ & $0^{0}_{0}$ & $0^{0}_{0}$ \\
Dome C & $0^{0}_{0}$ & $0^{0}_{0}$ & $0^{0}_{0}$ & $0^{0}_{0}$ & $0^{0}_{0}$ & $0^{0}_{0}$ \\
Dome F & $0^{0}_{0}$ & $0^{0}_{0}$ & $0^{0}_{0}$ & $0^{0}_{0}$ & $0^{0}_{0}$ & $0^{0}_{0}$ \\
ERB & $5^{7}_{4}$ & $7^{9}_{6}$ & $3^{5}_{3}$ & $0^{0}_{0}$ & $0^{1}_{0}$ & $5^{7}_{4}$ \\
FAIR & $3^{5}_{3}$ & $3^{4}_{2}$ & $12^{14}_{11}$ & $19^{22}_{18}$ & $10^{13}_{9}$ & $4^{6}_{3}$ \\
FUJI & $1^{2}_{0}$ & $2^{4}_{2}$ & $4^{6}_{3}$ & $10^{12}_{9}$ & $7^{9}_{6}$ & $2^{4}_{2}$ \\
GAM & $8^{10}_{7}$ & $7^{9}_{6}$ & $0^{1}_{0}$ & $0^{0}_{0}$ & $0^{1}_{0}$ & $1^{2}_{1}$ \\
GARS & $22^{24}_{21}$ & $21^{24}_{20}$ & $18^{20}_{16}$ & $14^{17}_{13}$ & $15^{17}_{14}$ & $21^{23}_{20}$ \\
GLT-S & $0^{0}_{0}$ & $0^{0}_{0}$ & $1^{2}_{0}$ & $1^{2}_{1}$ & $1^{3}_{1}$ & $0^{1}_{0}$ \\
HAN & $0^{1}_{0}$ & $1^{2}_{1}$ & $3^{5}_{2}$ & $7^{10}_{6}$ & $4^{5}_{3}$ & $0^{1}_{0}$ \\
HAY & $9^{11}_{7}$ & $10^{13}_{9}$ & $13^{15}_{11}$ & $11^{13}_{10}$ & $8^{10}_{7}$ & $9^{11}_{7}$ \\
HOP & $2^{4}_{2}$ & $1^{2}_{0}$ & $0^{1}_{0}$ & $12^{15}_{11}$ & $6^{8}_{5}$ & $1^{2}_{1}$ \\
JELM & $2^{3}_{1}$ & $3^{5}_{2}$ & $7^{9}_{6}$ & $6^{8}_{5}$ & $4^{6}_{3}$ & $2^{4}_{1}$ \\
KEN & $4^{6}_{3}$ & $9^{11}_{7}$ & $10^{12}_{9}$ & $11^{14}_{10}$ & $8^{10}_{7}$ & $15^{18}_{14}$ \\
KILI & $5^{7}_{4}$ & $9^{12}_{8}$ & $4^{6}_{3}$ & $0^{0}_{0}$ & $1^{2}_{0}$ & $11^{13}_{10}$ \\
KVNYS & $4^{5}_{3}$ & $5^{7}_{4}$ & $6^{8}_{5}$ & $13^{15}_{11}$ & $7^{9}_{6}$ & $8^{10}_{7}$ \\
LAS & $0^{0}_{0}$ & $0^{1}_{0}$ & $1^{3}_{1}$ & $1^{2}_{0}$ & $0^{1}_{0}$ & $0^{0}_{0}$ \\
LLA & $5^{7}_{4}$ & $2^{3}_{1}$ & $0^{0}_{0}$ & $0^{0}_{0}$ & $0^{0}_{0}$ & $0^{0}_{0}$ \\
LOS & $2^{3}_{1}$ & $2^{3}_{1}$ & $3^{4}_{2}$ & $9^{11}_{8}$ & $6^{8}_{5}$ & $2^{3}_{1}$ \\
NOB & $5^{7}_{4}$ & $9^{11}_{8}$ & $10^{12}_{9}$ & $19^{21}_{18}$ & $16^{18}_{14}$ & $10^{12}_{9}$ \\
NOR & $4^{6}_{3}$ & $6^{8}_{5}$ & $8^{10}_{7}$ & $16^{18}_{14}$ & $11^{13}_{9}$ & $4^{6}_{3}$ \\
NZ & $12^{14}_{10}$ & $9^{11}_{8}$ & $11^{13}_{10}$ & $10^{12}_{8}$ & $11^{14}_{10}$ & $12^{14}_{11}$ \\
ORG & $4^{6}_{3}$ & $8^{10}_{6}$ & $7^{9}_{6}$ & $1^{2}_{1}$ & $2^{4}_{2}$ & $5^{7}_{4}$ \\
OVRO & $3^{4}_{2}$ & $3^{5}_{2}$ & $4^{6}_{3}$ & $4^{5}_{3}$ & $1^{3}_{1}$ & $2^{3}_{1}$ \\
PAR & $0^{0}_{0}$ & $0^{0}_{0}$ & $0^{1}_{0}$ & $0^{0}_{0}$ & $0^{0}_{0}$ & $0^{0}_{0}$ \\
PIKE & $1^{2}_{0}$ & $1^{3}_{1}$ & $3^{5}_{3}$ & $8^{10}_{7}$ & $3^{5}_{3}$ & $0^{2}_{0}$ \\
SAN & $2^{3}_{1}$ & $1^{2}_{0}$ & $1^{2}_{0}$ & $3^{5}_{2}$ & $1^{3}_{1}$ & $0^{1}_{0}$ \\
SGO & $1^{2}_{0}$ & $0^{1}_{0}$ & $2^{3}_{1}$ & $1^{3}_{1}$ & $1^{3}_{1}$ & $0^{1}_{0}$ \\
SOC & $2^{3}_{1}$ & $2^{3}_{1}$ & $2^{4}_{2}$ & $11^{13}_{10}$ & $8^{10}_{6}$ & $1^{3}_{1}$ \\
SPX & $2^{4}_{2}$ & $3^{5}_{3}$ & $12^{14}_{10}$ & $12^{14}_{11}$ & $7^{9}_{5}$ & $3^{4}_{2}$ \\
SUF & $4^{5}_{3}$ & $6^{8}_{5}$ & $9^{11}_{8}$ & $1^{2}_{1}$ & $0^{2}_{0}$ & $3^{5}_{2}$ \\
YAN & $19^{21}_{17}$ & $20^{23}_{19}$ & $9^{11}_{8}$ & $4^{6}_{3}$ & $9^{11}_{8}$ & $12^{14}_{10}$ \\
YBG & $0^{0}_{0}$ & $0^{1}_{0}$ & $4^{6}_{3}$ & $16^{18}_{14}$ & $10^{12}_{9}$ & $0^{0}_{0}$ \\
\\
\enddata
\tablenotetext{}{\textbf{Note.} Sub/superscripts are 25 / 75\textsuperscript{th} percentiles, respectively.}
\end{deluxetable}
\label{table:clouds}

\begin{deluxetable*}{cccggcccccccc}[ht]
\tablewidth{90pt}
\tabletypesize{\footnotesize}
\tablecaption{Zenith Opacity at 230 GHz.}
\tablehead{\colhead{\textbf{site}} & \colhead{\textbf{Jan.}} & \colhead{\textbf{Feb.}} & \colhead{\textbf{Mar.}} & \colhead{\textbf{Apr.}} & \colhead{\textbf{May}} & \colhead{\textbf{Jun.}} & \colhead{\textbf{Jul.}} & \colhead{\textbf{Aug.}} & \colhead{\textbf{Sep.}} & \colhead{\textbf{Oct.}} & \colhead{\textbf{Nov.}} & \colhead{\textbf{Dec.}}} 
\startdata
ALMA/APEX & $0.13^{0.26}_{0.06}$ & $0.15^{0.31}_{0.07}$ & $0.08^{0.16}_{0.04}$ & $0.06^{0.11}_{0.03}$ & $0.04^{0.07}_{0.03}$ & $0.04^{0.06}_{0.03}$ & $0.03^{0.06}_{0.02}$ & $0.03^{0.05}_{0.03}$ & $0.04^{0.06}_{0.03}$ & $0.04^{0.06}_{0.03}$ & $0.04^{0.07}_{0.03}$ & $0.07^{0.15}_{0.04}$ \\
GLT & $0.16^{0.26}_{0.13}$ & $0.15^{0.25}_{0.12}$ & $0.15^{0.24}_{0.13}$ & $0.21^{0.31}_{0.16}$ & $0.37^{0.58}_{0.29}$ & $0.65^{0.94}_{0.53}$ & $0.84^{1.17}_{0.69}$ & $0.82^{1.27}_{0.62}$ & $0.53^{0.82}_{0.39}$ & $0.37^{0.60}_{0.24}$ & $0.24^{0.40}_{0.17}$ & $0.17^{0.30}_{0.13}$ \\
IRAM-30m & $0.14^{0.24}_{0.08}$ & $0.13^{0.24}_{0.07}$ & $0.15^{0.29}_{0.08}$ & $0.20^{0.34}_{0.12}$ & $0.24^{0.37}_{0.14}$ & $0.30^{0.43}_{0.20}$ & $0.26^{0.40}_{0.16}$ & $0.33^{0.49}_{0.21}$ & $0.29^{0.45}_{0.19}$ & $0.26^{0.42}_{0.15}$ & $0.19^{0.36}_{0.11}$ & $0.15^{0.27}_{0.08}$ \\
JCMT/SMA & $0.06^{0.12}_{0.04}$ & $0.08^{0.17}_{0.04}$ & $0.08^{0.20}_{0.04}$ & $0.08^{0.16}_{0.05}$ & $0.08^{0.16}_{0.05}$ & $0.09^{0.15}_{0.06}$ & $0.11^{0.17}_{0.06}$ & $0.09^{0.21}_{0.06}$ & $0.08^{0.17}_{0.05}$ & $0.11^{0.22}_{0.06}$ & $0.09^{0.21}_{0.05}$ & $0.07^{0.17}_{0.04}$ \\
KP & $0.21^{0.40}_{0.13}$ & $0.21^{0.36}_{0.13}$ & $0.24^{0.37}_{0.15}$ & $0.25^{0.37}_{0.16}$ & $0.29^{0.43}_{0.20}$ & $0.45^{0.78}_{0.26}$ & $1.47^{1.77}_{1.11}$ & $1.31^{1.66}_{0.97}$ & $0.92^{1.31}_{0.54}$ & $0.40^{0.65}_{0.25}$ & $0.26^{0.42}_{0.16}$ & $0.24^{0.40}_{0.14}$ \\
LMT & $0.09^{0.15}_{0.05}$ & $0.09^{0.16}_{0.05}$ & $0.13^{0.21}_{0.06}$ & $0.19^{0.33}_{0.12}$ & $0.25^{0.45}_{0.15}$ & $0.40^{0.63}_{0.20}$ & $0.37^{0.56}_{0.22}$ & $0.39^{0.58}_{0.26}$ & $0.49^{0.69}_{0.31}$ & $0.26^{0.46}_{0.11}$ & $0.14^{0.27}_{0.07}$ & $0.09^{0.16}_{0.05}$ \\
NOEMA & $0.16^{0.29}_{0.09}$ & $0.16^{0.30}_{0.09}$ & $0.19^{0.33}_{0.12}$ & $0.27^{0.45}_{0.17}$ & $0.37^{0.62}_{0.22}$ & $0.46^{0.71}_{0.31}$ & $0.48^{0.69}_{0.31}$ & $0.45^{0.67}_{0.29}$ & $0.39^{0.60}_{0.24}$ & $0.29^{0.51}_{0.16}$ & $0.25^{0.45}_{0.14}$ & $0.17^{0.31}_{0.09}$ \\
SMT & $0.12^{0.20}_{0.07}$ & $0.11^{0.19}_{0.06}$ & $0.13^{0.20}_{0.07}$ & $0.15^{0.22}_{0.09}$ & $0.17^{0.26}_{0.12}$ & $0.30^{0.51}_{0.16}$ & $0.82^{1.05}_{0.60}$ & $0.73^{0.99}_{0.51}$ & $0.48^{0.73}_{0.29}$ & $0.19^{0.34}_{0.12}$ & $0.13^{0.22}_{0.07}$ & $0.13^{0.21}_{0.07}$ \\
SPT & $0.06^{0.07}_{0.05}$ & $0.05^{0.06}_{0.05}$ & $0.05^{0.05}_{0.04}$ & $0.04^{0.05}_{0.04}$ & $0.04^{0.05}_{0.04}$ & $0.04^{0.04}_{0.03}$ & $0.04^{0.04}_{0.03}$ & $0.04^{0.04}_{0.03}$ & $0.04^{0.04}_{0.03}$ & $0.04^{0.05}_{0.04}$ & $0.04^{0.05}_{0.04}$ & $0.06^{0.07}_{0.05}$ \\
\\
\hline
BAJA & $0.13^{0.25}_{0.07}$ & $0.11^{0.20}_{0.06}$ & $0.12^{0.21}_{0.07}$ & $0.13^{0.21}_{0.07}$ & $0.14^{0.21}_{0.09}$ & $0.19^{0.39}_{0.12}$ & $0.67^{0.94}_{0.40}$ & $0.56^{0.82}_{0.31}$ & $0.34^{0.62}_{0.16}$ & $0.16^{0.27}_{0.09}$ & $0.13^{0.21}_{0.07}$ & $0.13^{0.22}_{0.07}$ \\
BAN & $0.24^{0.42}_{0.14}$ & $0.20^{0.37}_{0.12}$ & $0.27^{0.45}_{0.16}$ & $0.30^{0.49}_{0.20}$ & $0.43^{0.65}_{0.29}$ & $0.62^{0.89}_{0.45}$ & $0.69^{0.99}_{0.51}$ & $0.65^{0.94}_{0.49}$ & $0.48^{0.73}_{0.33}$ & $0.36^{0.58}_{0.22}$ & $0.27^{0.46}_{0.16}$ & $0.21^{0.36}_{0.13}$ \\
BAR & $0.11^{0.17}_{0.05}$ & $0.08^{0.14}_{0.05}$ & $0.11^{0.16}_{0.06}$ & $0.12^{0.19}_{0.07}$ & $0.16^{0.25}_{0.11}$ & $0.19^{0.29}_{0.12}$ & $0.29^{0.48}_{0.16}$ & $0.21^{0.37}_{0.13}$ & $0.17^{0.27}_{0.11}$ & $0.14^{0.21}_{0.08}$ & $0.11^{0.16}_{0.06}$ & $0.09^{0.16}_{0.05}$ \\
BGA & $0.15^{0.26}_{0.09}$ & $0.16^{0.27}_{0.11}$ & $0.19^{0.33}_{0.11}$ & $0.25^{0.43}_{0.14}$ & $0.37^{0.60}_{0.24}$ & $0.46^{0.73}_{0.29}$ & $0.42^{0.67}_{0.27}$ & $0.39^{0.58}_{0.26}$ & $0.36^{0.58}_{0.21}$ & $0.26^{0.45}_{0.13}$ & $0.21^{0.33}_{0.12}$ & $0.15^{0.26}_{0.08}$ \\
BGK & $0.58^{0.94}_{0.34}$ & $0.54^{0.89}_{0.31}$ & $0.51^{0.87}_{0.29}$ & $0.54^{0.94}_{0.30}$ & $0.62^{1.08}_{0.37}$ & $0.76^{1.20}_{0.51}$ & $0.89^{1.39}_{0.65}$ & $0.92^{1.47}_{0.63}$ & $0.89^{1.43}_{0.54}$ & $0.73^{1.17}_{0.43}$ & $0.60^{1.02}_{0.34}$ & $0.53^{0.92}_{0.29}$ \\
BLDR & $0.21^{0.30}_{0.13}$ & $0.20^{0.33}_{0.14}$ & $0.24^{0.36}_{0.16}$ & $0.31^{0.48}_{0.24}$ & $0.46^{0.63}_{0.33}$ & $0.54^{0.76}_{0.39}$ & $0.84^{1.11}_{0.62}$ & $0.73^{0.99}_{0.54}$ & $0.53^{0.76}_{0.37}$ & $0.36^{0.51}_{0.25}$ & $0.24^{0.34}_{0.15}$ & $0.20^{0.30}_{0.13}$ \\
BMAC & $0.63^{0.99}_{0.40}$ & $0.62^{1.02}_{0.36}$ & $0.49^{0.76}_{0.29}$ & $0.30^{0.53}_{0.16}$ & $0.22^{0.39}_{0.13}$ & $0.12^{0.24}_{0.07}$ & $0.13^{0.22}_{0.07}$ & $0.16^{0.29}_{0.08}$ & $0.21^{0.39}_{0.12}$ & $0.30^{0.51}_{0.17}$ & $0.37^{0.60}_{0.22}$ & $0.49^{0.80}_{0.30}$ \\
BOL & $0.39^{0.51}_{0.27}$ & $0.42^{0.54}_{0.29}$ & $0.34^{0.49}_{0.22}$ & $0.24^{0.40}_{0.14}$ & $0.15^{0.29}_{0.08}$ & $0.12^{0.24}_{0.05}$ & $0.09^{0.20}_{0.05}$ & $0.11^{0.22}_{0.05}$ & $0.17^{0.34}_{0.08}$ & $0.27^{0.43}_{0.14}$ & $0.27^{0.43}_{0.16}$ & $0.37^{0.51}_{0.25}$ \\
BRZ & $0.71^{1.35}_{0.34}$ & $0.63^{1.14}_{0.36}$ & $0.97^{1.47}_{0.54}$ & $0.63^{1.14}_{0.31}$ & $0.43^{0.76}_{0.21}$ & $0.33^{0.58}_{0.15}$ & $0.24^{0.46}_{0.11}$ & $0.22^{0.45}_{0.09}$ & $0.36^{0.69}_{0.13}$ & $0.71^{1.17}_{0.36}$ & $0.94^{1.47}_{0.49}$ & $0.94^{1.51}_{0.54}$ \\
CAS & $0.80^{1.27}_{0.53}$ & $0.76^{1.24}_{0.48}$ & $0.71^{1.17}_{0.46}$ & $0.60^{1.02}_{0.37}$ & $0.49^{0.82}_{0.33}$ & $0.48^{0.78}_{0.31}$ & $0.45^{0.73}_{0.27}$ & $0.46^{0.76}_{0.29}$ & $0.48^{0.78}_{0.30}$ & $0.54^{0.89}_{0.34}$ & $0.63^{1.05}_{0.42}$ & $0.73^{1.14}_{0.48}$ \\
CAT & $0.31^{0.53}_{0.20}$ & $0.34^{0.58}_{0.20}$ & $0.31^{0.54}_{0.19}$ & $0.29^{0.54}_{0.15}$ & $0.31^{0.62}_{0.14}$ & $0.30^{0.54}_{0.15}$ & $0.26^{0.48}_{0.13}$ & $0.29^{0.54}_{0.13}$ & $0.24^{0.48}_{0.13}$ & $0.27^{0.49}_{0.15}$ & $0.29^{0.51}_{0.17}$ & $0.31^{0.58}_{0.20}$ \\
CNI & $0.16^{0.27}_{0.09}$ & $0.13^{0.25}_{0.08}$ & $0.16^{0.27}_{0.08}$ & $0.16^{0.29}_{0.11}$ & $0.19^{0.30}_{0.12}$ & $0.22^{0.37}_{0.15}$ & $0.26^{0.46}_{0.15}$ & $0.39^{0.60}_{0.20}$ & $0.33^{0.54}_{0.20}$ & $0.30^{0.53}_{0.17}$ & $0.22^{0.42}_{0.12}$ & $0.17^{0.34}_{0.09}$ \\
Dome~A & $0.03^{0.04}_{0.03}$ & $0.03^{0.03}_{0.03}$ & $0.03^{0.03}_{0.02}$ & $0.02^{0.03}_{0.02}$ & $0.02^{0.03}_{0.02}$ & $0.02^{0.03}_{0.02}$ & $0.02^{0.03}_{0.02}$ & $0.02^{0.02}_{0.02}$ & $0.02^{0.02}_{0.02}$ & $0.02^{0.03}_{0.02}$ & $0.03^{0.03}_{0.03}$ & $0.03^{0.04}_{0.03}$ \\
Dome~C & $0.05^{0.06}_{0.04}$ & $0.04^{0.05}_{0.04}$ & $0.04^{0.04}_{0.03}$ & $0.03^{0.04}_{0.03}$ & $0.03^{0.04}_{0.03}$ & $0.03^{0.03}_{0.03}$ & $0.03^{0.03}_{0.03}$ & $0.03^{0.04}_{0.03}$ & $0.03^{0.03}_{0.03}$ & $0.03^{0.04}_{0.03}$ & $0.04^{0.04}_{0.03}$ & $0.05^{0.05}_{0.04}$ \\
Dome~F & $0.04^{0.05}_{0.04}$ & $0.04^{0.04}_{0.03}$ & $0.03^{0.03}_{0.03}$ & $0.03^{0.03}_{0.03}$ & $0.03^{0.03}_{0.03}$ & $0.03^{0.03}_{0.02}$ & $0.03^{0.03}_{0.02}$ & $0.03^{0.03}_{0.02}$ & $0.03^{0.03}_{0.02}$ & $0.03^{0.03}_{0.03}$ & $0.03^{0.04}_{0.03}$ & $0.04^{0.04}_{0.03}$ \\
ERB & $0.24^{0.40}_{0.15}$ & $0.27^{0.43}_{0.15}$ & $0.31^{0.51}_{0.20}$ & $0.42^{0.62}_{0.27}$ & $0.54^{0.73}_{0.39}$ & $0.48^{0.65}_{0.33}$ & $0.49^{0.71}_{0.33}$ & $0.46^{0.63}_{0.31}$ & $0.43^{0.58}_{0.30}$ & $0.45^{0.67}_{0.29}$ & $0.33^{0.54}_{0.19}$ & $0.26^{0.46}_{0.15}$ \\
FAIR & $0.21^{0.34}_{0.11}$ & $0.20^{0.31}_{0.13}$ & $0.17^{0.29}_{0.11}$ & $0.27^{0.42}_{0.19}$ & $0.49^{0.78}_{0.36}$ & $0.80^{1.14}_{0.60}$ & $1.02^{1.43}_{0.78}$ & $0.87^{1.27}_{0.65}$ & $0.56^{0.84}_{0.39}$ & $0.36^{0.53}_{0.24}$ & $0.20^{0.36}_{0.13}$ & $0.22^{0.36}_{0.14}$ \\
FUJI & $0.04^{0.08}_{0.03}$ & $0.05^{0.13}_{0.03}$ & $0.06^{0.16}_{0.03}$ & $0.08^{0.22}_{0.04}$ & $0.13^{0.34}_{0.05}$ & $0.29^{0.63}_{0.12}$ & $0.49^{0.80}_{0.29}$ & $0.49^{0.76}_{0.29}$ & $0.34^{0.63}_{0.13}$ & $0.13^{0.39}_{0.05}$ & $0.07^{0.19}_{0.03}$ & $0.04^{0.11}_{0.03}$ \\
GAM & $0.94^{1.35}_{0.60}$ & $1.02^{1.51}_{0.60}$ & $0.84^{1.24}_{0.53}$ & $0.56^{0.82}_{0.34}$ & $0.33^{0.48}_{0.21}$ & $0.19^{0.29}_{0.11}$ & $0.19^{0.30}_{0.11}$ & $0.20^{0.33}_{0.12}$ & $0.30^{0.51}_{0.14}$ & $0.42^{0.63}_{0.21}$ & $0.48^{0.73}_{0.25}$ & $0.62^{0.94}_{0.39}$ \\
GARS & $0.71^{1.14}_{0.51}$ & $0.76^{1.20}_{0.51}$ & $0.78^{1.24}_{0.48}$ & $0.60^{1.05}_{0.34}$ & $0.58^{1.05}_{0.33}$ & $0.53^{0.94}_{0.27}$ & $0.40^{0.82}_{0.21}$ & $0.40^{0.82}_{0.21}$ & $0.48^{0.94}_{0.25}$ & $0.58^{1.05}_{0.34}$ & $0.62^{1.05}_{0.42}$ & $0.67^{1.05}_{0.48}$ \\
GLT-S & $0.06^{0.09}_{0.05}$ & $0.06^{0.08}_{0.05}$ & $0.05^{0.08}_{0.04}$ & $0.06^{0.11}_{0.05}$ & $0.09^{0.15}_{0.07}$ & $0.13^{0.20}_{0.09}$ & $0.16^{0.22}_{0.12}$ & $0.15^{0.24}_{0.12}$ & $0.13^{0.20}_{0.08}$ & $0.08^{0.14}_{0.06}$ & $0.06^{0.11}_{0.05}$ & $0.05^{0.08}_{0.04}$ \\
HAN & $0.07^{0.12}_{0.05}$ & $0.11^{0.15}_{0.07}$ & $0.12^{0.19}_{0.08}$ & $0.17^{0.25}_{0.12}$ & $0.22^{0.31}_{0.15}$ & $0.30^{0.43}_{0.21}$ & $0.53^{0.69}_{0.37}$ & $0.56^{0.71}_{0.37}$ & $0.26^{0.45}_{0.15}$ & $0.12^{0.16}_{0.08}$ & $0.08^{0.13}_{0.06}$ & $0.07^{0.11}_{0.05}$ \\
HAY & $0.42^{0.73}_{0.22}$ & $0.45^{0.82}_{0.26}$ & $0.54^{0.99}_{0.30}$ & $0.78^{1.35}_{0.48}$ & $1.35^{2.21}_{0.89}$ & $1.77^{2.66}_{1.20}$ & $2.21^{3.00}_{1.56}$ & $2.04^{2.81}_{1.43}$ & $1.51^{2.41}_{0.94}$ & $1.05^{1.83}_{0.65}$ & $0.65^{1.14}_{0.39}$ & $0.53^{0.97}_{0.31}$ \\
HOP & $0.17^{0.31}_{0.11}$ & $0.17^{0.30}_{0.11}$ & $0.19^{0.30}_{0.12}$ & $0.21^{0.31}_{0.14}$ & $0.24^{0.36}_{0.16}$ & $0.42^{0.73}_{0.25}$ & $1.27^{1.61}_{0.99}$ & $1.14^{1.51}_{0.87}$ & $0.80^{1.17}_{0.49}$ & $0.33^{0.56}_{0.20}$ & $0.21^{0.34}_{0.13}$ & $0.19^{0.33}_{0.11}$ \\
JELM & $0.19^{0.27}_{0.11}$ & $0.17^{0.30}_{0.12}$ & $0.21^{0.31}_{0.14}$ & $0.27^{0.42}_{0.20}$ & $0.39^{0.54}_{0.29}$ & $0.45^{0.60}_{0.31}$ & $0.65^{0.89}_{0.46}$ & $0.58^{0.80}_{0.42}$ & $0.43^{0.62}_{0.29}$ & $0.30^{0.45}_{0.21}$ & $0.20^{0.30}_{0.13}$ & $0.17^{0.27}_{0.11}$ \\
KEN & $0.16^{0.31}_{0.09}$ & $0.17^{0.34}_{0.09}$ & $0.22^{0.48}_{0.11}$ & $0.46^{0.71}_{0.26}$ & $0.39^{0.60}_{0.24}$ & $0.33^{0.54}_{0.20}$ & $0.34^{0.53}_{0.19}$ & $0.31^{0.53}_{0.16}$ & $0.26^{0.45}_{0.16}$ & $0.34^{0.54}_{0.21}$ & $0.43^{0.63}_{0.27}$ & $0.30^{0.51}_{0.17}$ \\
KILI & $0.19^{0.37}_{0.09}$ & $0.17^{0.39}_{0.09}$ & $0.26^{0.49}_{0.13}$ & $0.37^{0.58}_{0.21}$ & $0.26^{0.42}_{0.15}$ & $0.15^{0.24}_{0.08}$ & $0.09^{0.17}_{0.06}$ & $0.12^{0.21}_{0.06}$ & $0.13^{0.21}_{0.08}$ & $0.17^{0.33}_{0.09}$ & $0.31^{0.51}_{0.17}$ & $0.31^{0.51}_{0.17}$ \\
KVNYS & $0.24^{0.40}_{0.16}$ & $0.29^{0.49}_{0.17}$ & $0.39^{0.63}_{0.24}$ & $0.65^{1.05}_{0.42}$ & $1.02^{1.56}_{0.67}$ & $1.97^{2.53}_{1.43}$ & $3.22^{3.91}_{2.41}$ & $3.00^{3.91}_{2.21}$ & $1.56^{2.30}_{1.05}$ & $0.80^{1.24}_{0.53}$ & $0.56^{0.99}_{0.31}$ & $0.29^{0.53}_{0.17}$ \\
LAS & $0.29^{0.46}_{0.17}$ & $0.27^{0.42}_{0.17}$ & $0.22^{0.34}_{0.15}$ & $0.19^{0.29}_{0.13}$ & $0.20^{0.30}_{0.13}$ & $0.16^{0.26}_{0.11}$ & $0.14^{0.22}_{0.08}$ & $0.14^{0.22}_{0.09}$ & $0.14^{0.22}_{0.09}$ & $0.15^{0.24}_{0.11}$ & $0.17^{0.25}_{0.12}$ & $0.20^{0.29}_{0.13}$ \\
LLA & $0.22^{0.40}_{0.11}$ & $0.26^{0.46}_{0.12}$ & $0.13^{0.22}_{0.06}$ & $0.07^{0.13}_{0.04}$ & $0.05^{0.08}_{0.03}$ & $0.04^{0.06}_{0.03}$ & $0.04^{0.06}_{0.03}$ & $0.04^{0.06}_{0.03}$ & $0.05^{0.07}_{0.03}$ & $0.04^{0.07}_{0.03}$ & $0.06^{0.09}_{0.04}$ & $0.12^{0.25}_{0.06}$ \\
LOS & $0.24^{0.36}_{0.16}$ & $0.26^{0.36}_{0.17}$ & $0.27^{0.39}_{0.19}$ & $0.34^{0.46}_{0.24}$ & $0.42^{0.58}_{0.30}$ & $0.56^{0.87}_{0.34}$ & $1.17^{1.47}_{0.92}$ & $1.08^{1.35}_{0.80}$ & $0.80^{1.11}_{0.53}$ & $0.43^{0.63}_{0.29}$ & $0.27^{0.40}_{0.17}$ & $0.25^{0.37}_{0.16}$ \\
NOB & $0.24^{0.40}_{0.16}$ & $0.25^{0.48}_{0.16}$ & $0.31^{0.63}_{0.19}$ & $0.46^{0.89}_{0.27}$ & $0.71^{1.24}_{0.42}$ & $1.35^{2.12}_{0.89}$ & $2.04^{2.66}_{1.51}$ & $2.12^{2.81}_{1.61}$ & $1.51^{2.30}_{0.89}$ & $0.73^{1.43}_{0.43}$ & $0.43^{0.80}_{0.26}$ & $0.29^{0.49}_{0.19}$ \\
NOR & $0.08^{0.17}_{0.04}$ & $0.08^{0.21}_{0.04}$ & $0.09^{0.26}_{0.05}$ & $0.14^{0.37}_{0.06}$ & $0.21^{0.54}_{0.09}$ & $0.48^{0.99}_{0.22}$ & $0.84^{1.31}_{0.51}$ & $0.87^{1.35}_{0.53}$ & $0.53^{1.05}_{0.17}$ & $0.16^{0.53}_{0.06}$ & $0.12^{0.29}_{0.05}$ & $0.09^{0.21}_{0.05}$ \\
NZ & $0.43^{0.87}_{0.24}$ & $0.40^{0.76}_{0.24}$ & $0.34^{0.71}_{0.19}$ & $0.34^{0.69}_{0.17}$ & $0.30^{0.58}_{0.17}$ & $0.24^{0.48}_{0.13}$ & $0.24^{0.45}_{0.12}$ & $0.29^{0.53}_{0.14}$ & $0.30^{0.56}_{0.15}$ & $0.34^{0.62}_{0.17}$ & $0.37^{0.69}_{0.21}$ & $0.43^{0.82}_{0.24}$ \\
ORG & $0.24^{0.42}_{0.13}$ & $0.24^{0.40}_{0.15}$ & $0.27^{0.49}_{0.17}$ & $0.30^{0.49}_{0.20}$ & $0.42^{0.62}_{0.27}$ & $0.46^{0.65}_{0.33}$ & $0.53^{0.73}_{0.37}$ & $0.51^{0.69}_{0.37}$ & $0.42^{0.58}_{0.29}$ & $0.36^{0.51}_{0.22}$ & $0.27^{0.45}_{0.15}$ & $0.24^{0.43}_{0.13}$ \\
OVRO & $0.42^{0.69}_{0.24}$ & $0.42^{0.56}_{0.31}$ & $0.49^{0.65}_{0.33}$ & $0.51^{0.69}_{0.39}$ & $0.67^{0.89}_{0.48}$ & $0.67^{0.94}_{0.48}$ & $0.97^{1.43}_{0.60}$ & $0.76^{1.17}_{0.51}$ & $0.67^{0.97}_{0.48}$ & $0.56^{0.80}_{0.40}$ & $0.45^{0.65}_{0.31}$ & $0.40^{0.62}_{0.26}$ \\
PAR & $0.20^{0.33}_{0.12}$ & $0.21^{0.39}_{0.14}$ & $0.16^{0.26}_{0.12}$ & $0.14^{0.22}_{0.09}$ & $0.13^{0.21}_{0.09}$ & $0.11^{0.17}_{0.07}$ & $0.09^{0.14}_{0.06}$ & $0.09^{0.15}_{0.07}$ & $0.09^{0.15}_{0.06}$ & $0.11^{0.15}_{0.07}$ & $0.11^{0.15}_{0.08}$ & $0.14^{0.21}_{0.09}$ \\
PIKE & $0.07^{0.12}_{0.05}$ & $0.07^{0.11}_{0.04}$ & $0.08^{0.13}_{0.05}$ & $0.12^{0.19}_{0.07}$ & $0.16^{0.26}_{0.11}$ & $0.20^{0.31}_{0.13}$ & $0.36^{0.54}_{0.22}$ & $0.31^{0.48}_{0.21}$ & $0.20^{0.33}_{0.12}$ & $0.12^{0.19}_{0.07}$ & $0.08^{0.12}_{0.05}$ & $0.07^{0.11}_{0.05}$ \\
SAN & $0.15^{0.27}_{0.08}$ & $0.13^{0.22}_{0.07}$ & $0.15^{0.25}_{0.08}$ & $0.16^{0.26}_{0.11}$ & $0.21^{0.33}_{0.13}$ & $0.21^{0.36}_{0.13}$ & $0.63^{0.92}_{0.27}$ & $0.48^{0.78}_{0.26}$ & $0.31^{0.54}_{0.17}$ & $0.20^{0.33}_{0.12}$ & $0.15^{0.25}_{0.09}$ & $0.15^{0.26}_{0.08}$ \\
SGO & $0.16^{0.26}_{0.11}$ & $0.15^{0.25}_{0.09}$ & $0.13^{0.20}_{0.08}$ & $0.12^{0.19}_{0.07}$ & $0.13^{0.21}_{0.07}$ & $0.11^{0.19}_{0.06}$ & $0.11^{0.17}_{0.06}$ & $0.11^{0.19}_{0.06}$ & $0.11^{0.17}_{0.06}$ & $0.12^{0.19}_{0.07}$ & $0.12^{0.19}_{0.08}$ & $0.13^{0.20}_{0.08}$ \\
SOC & $0.24^{0.37}_{0.15}$ & $0.25^{0.37}_{0.16}$ & $0.27^{0.39}_{0.19}$ & $0.31^{0.45}_{0.22}$ & $0.39^{0.54}_{0.26}$ & $0.56^{0.89}_{0.34}$ & $1.31^{1.66}_{1.05}$ & $1.24^{1.56}_{0.94}$ & $0.94^{1.27}_{0.62}$ & $0.45^{0.69}_{0.30}$ & $0.29^{0.43}_{0.17}$ & $0.26^{0.42}_{0.16}$ \\
SPX & $0.09^{0.17}_{0.05}$ & $0.09^{0.17}_{0.05}$ & $0.11^{0.19}_{0.06}$ & $0.15^{0.27}_{0.08}$ & $0.24^{0.40}_{0.13}$ & $0.30^{0.49}_{0.16}$ & $0.30^{0.53}_{0.16}$ & $0.29^{0.51}_{0.16}$ & $0.21^{0.39}_{0.12}$ & $0.15^{0.27}_{0.08}$ & $0.14^{0.25}_{0.07}$ & $0.09^{0.19}_{0.06}$ \\
SUF & $0.22^{0.37}_{0.13}$ & $0.25^{0.43}_{0.13}$ & $0.37^{0.60}_{0.22}$ & $0.46^{0.71}_{0.29}$ & $0.62^{0.89}_{0.43}$ & $0.65^{0.87}_{0.49}$ & $0.62^{0.78}_{0.48}$ & $0.54^{0.73}_{0.40}$ & $0.42^{0.54}_{0.30}$ & $0.36^{0.54}_{0.22}$ & $0.29^{0.45}_{0.16}$ & $0.20^{0.36}_{0.11}$ \\
YAN & $0.56^{0.67}_{0.42}$ & $0.60^{0.71}_{0.48}$ & $0.60^{0.71}_{0.46}$ & $0.49^{0.63}_{0.34}$ & $0.37^{0.53}_{0.24}$ & $0.29^{0.43}_{0.19}$ & $0.24^{0.39}_{0.14}$ & $0.27^{0.42}_{0.16}$ & $0.37^{0.51}_{0.24}$ & $0.43^{0.56}_{0.29}$ & $0.45^{0.60}_{0.29}$ & $0.53^{0.65}_{0.39}$ \\
YBG & $0.07^{0.09}_{0.05}$ & $0.08^{0.12}_{0.06}$ & $0.12^{0.19}_{0.08}$ & $0.17^{0.26}_{0.12}$ & $0.29^{0.43}_{0.20}$ & $0.58^{0.80}_{0.36}$ & $0.82^{1.02}_{0.71}$ & $0.78^{0.94}_{0.63}$ & $0.63^{0.78}_{0.45}$ & $0.20^{0.33}_{0.14}$ & $0.09^{0.13}_{0.07}$ & $0.07^{0.09}_{0.06}$ \\
\\
\enddata
\tablenotetext{}{\textbf{Note.} Sub/superscripts are 25 / 75\textsuperscript{th} percentiles, respectively.}
\end{deluxetable*}
\label{table:tau}

The opacities we report agree reasonably well with published field measurements.  At the Greenland Summit site, a multiyear 225~GHz radiometer campaign beginning in 2010 found that the median opacity during the winter months was about 0.06~\citep{Matsushita2017}.  We find a similar median opacity: 0.05 to 0.06.  The median in July is also similar: 0.16--0.19 in the \cite{Matsushita2017} field measurement and 0.16 for our retrieval.

At the HAN site, 220~GHz tipping radiometer measurements beginning at the end of year 1999 until the middle of 2001 have been published~\citep{Ananthasubramanian2002}.  The monthly median opacities reported are $0.07^{0.09}_{0.05}$ in January and $0.37^{0.46}_{0.26}$ in July.  For 2009--2019, we calculate opacities of $0.07^{0.12}_{0.05}$ and $0.53^{0.69}_{0.37}$ in those months.  The agreement in wintertime is excellent.  In summer, additional field measurement data is needed to constrain the level of overlap between the experimental and model distributions.

The wintertime conditions at FUJI are very good and agree with previous 220~GHz radiometer measurements done 1994--95~\citep{Sekimoto1996}.  In December of 1994, an opacity of about 0.04 was measured, which matches exactly with the $0.04^{0.11}_{0.03}$ range from our calculation.  The agreement in March is also within the statistical variation: 0.11 observed compared with $0.06^{0.16}_{0.03}$ from our 10 year average. Mt. Fuji has good submillimeter conditions in winter.

Finally, our opacity calculations agree with the 225~GHz tipping radiometer measurements for 2001 September to November, reported by \cite{Marvil2006} for the BAR site.  The median measured opacity was 0.11 over that time frame, which is similar to the 0.11 to 0.17 10 year median values we calculate for the same months.

\subsection{Applicability to 345~GHz and Higher Frequencies}
\begin{figure}[ht]
\centering
  \includegraphics[width=\columnwidth]{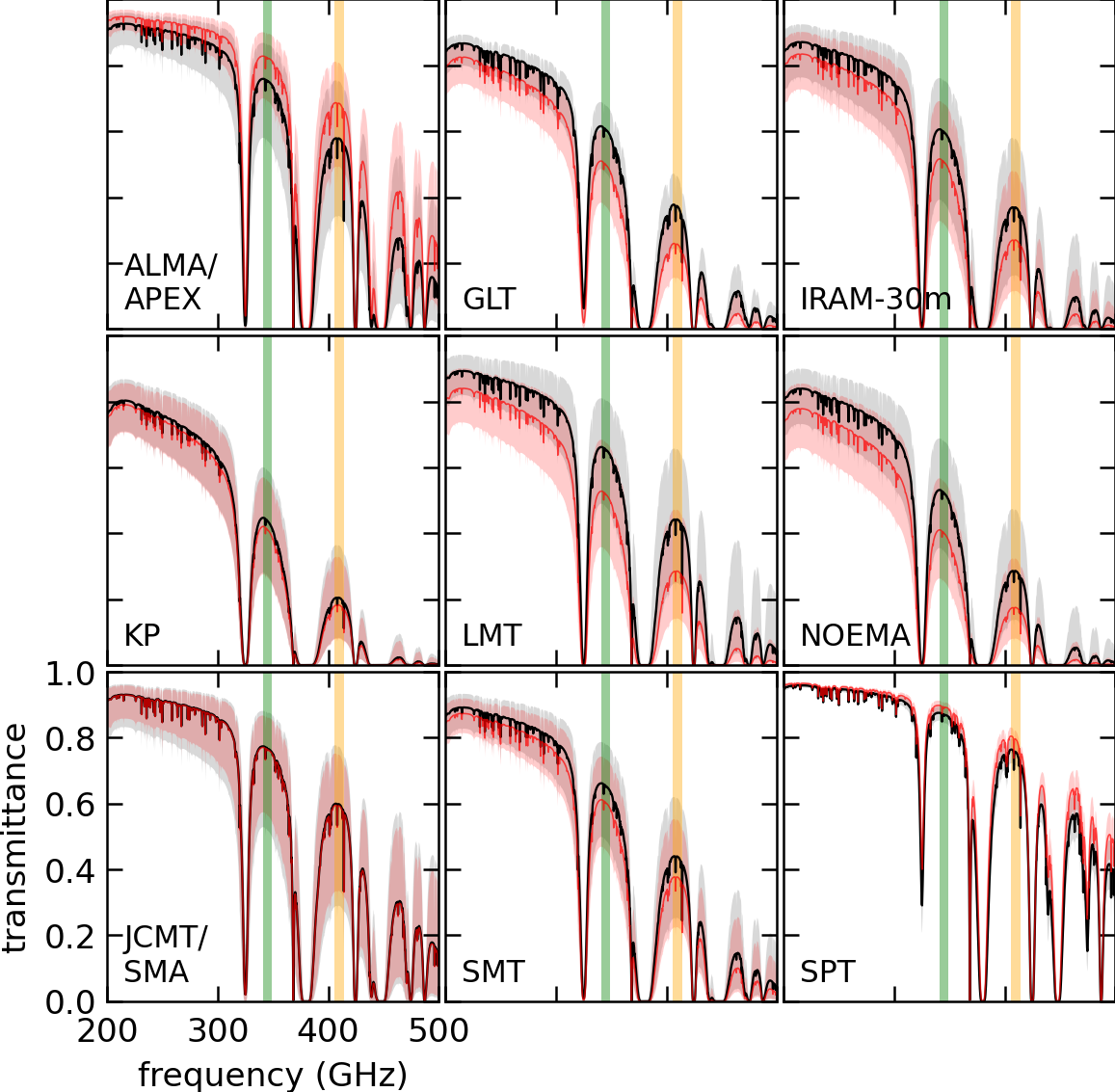}
  \caption{Zenith transmittance spectra at $200-500$\,GHz for existing EHT sites in March (black) and April (red).  The shaded region brackets the interquartile range, and 8~GHz-wide high-frequency bands centered at 345 (green) and 410~GHz (yellow) are marked with vertical lines.}
  \label{fig:spectraOLD}
\end{figure}

The highest-frequency Fourier components ($\vec{u}$) sampled during the 230~GHz observation in 2017 of the \m87 source were measured on the baseline between Hawaii and Spain~\citep{EHT2} and corresponded to 25\,$\mu$as instrumental angular resolution defined by the fringe spacing $\lambda/D$, where $\lambda$ is the observing wavelength and $D$ is the projected baseline length to the source.  There is interest in operating at 345~GHz or potentially higher frequencies at the stations with suitable conditions, which could improve the nominal resolution by 50\% or more.  Most of the 2021 EHT sites are capable of 345~GHz operation during at least part of the year, and the transmittances ($Tr = - \ln{\tau}$) of each existing site during March and April are plotted in Fig.~\ref{fig:spectraOLD}.  Although the transmittance threshold for 345~GHz capability depends on the telescope sensitivity and other details like the atmospheric coherence, we can make a reasonable estimate of the transmittance that is required.    Consider the case of 0.1~Jy correlated flux density at 45$^\circ$ elevation with two-bit digital efficiency, and a 345~GHz zenith transmittance of 0.6 (opacity of 0.5) and sky-brightness temperature of 150~K.  Under those conditions, two stations with 10~m diameter antennas having 64~$\mu$m effective surface-accuracy apertures and 100~K receiver temperatures integrating for 100~s at 8~GHz fringe-finding bandwidth would achieve an S/N 3 detection.  Stronger detections would be achieved on baselines to large apertures like ALMA or NOEMA.

\begin{figure*}[ht]
\centering
  \includegraphics[width=1.5\columnwidth]{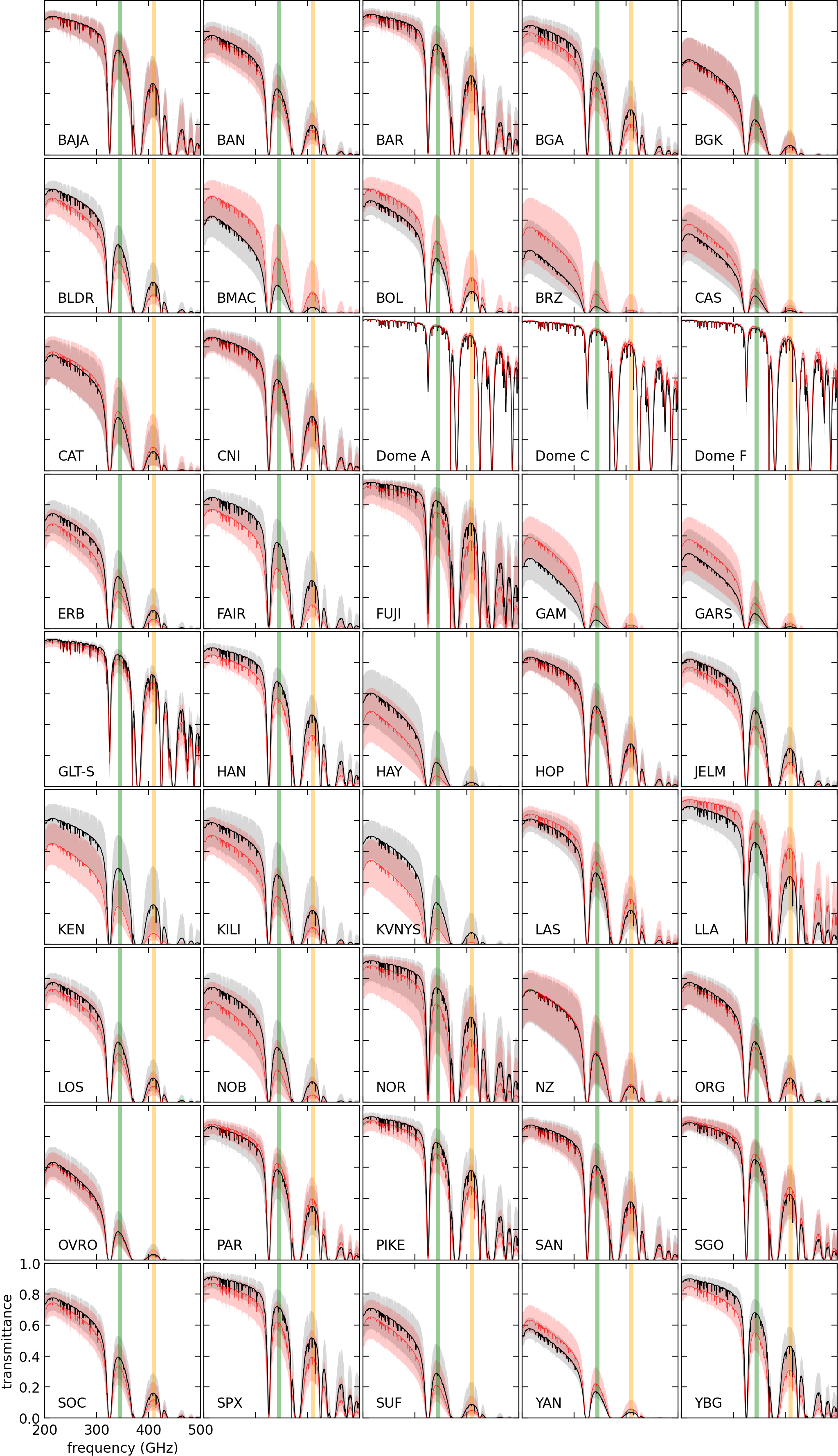}
  \caption{Zenith transmittance spectra at $200-500$\,GHz for candidate new EHT sites for March (black) and April (red).  Shaded spectra are the interquartile range, and high-frequency 8~GHz-wide high-frequency bands centered at 345 and 410~GHz are marked with vertical lines.}
  \label{fig:spectraNEW}
\end{figure*}

The existing sites in Chile, Hawaii, and the South Pole have median March 345~GHz transmittance of 0.75 or greater.  The South pole, in particular, has excellent 345~GHz transmittance even at the 25\textsuperscript{th} percentile.  The SPT site also has relatively good transmittance at 410~GHz (above 70\%).  Zenith transmission spectra for the new sites are plotted in Fig.~\ref{fig:spectraNEW}.  Of these, Dome~A , Dome~C, and Dome~F have high-frequency conditions that are similar to the SPT site.  The FUJI, GLT-S, HAN, LLA, and PIKE sites are similar to the ALMA/APEX and JCMT/SMA sites: more opaque than Antarctica and more variable.  BAJA, BAR, NOR, SGO, SPX, and YBG have comparable 345~GHz transmittance to the LMT site and would probably be viable during a useful fraction of time.

The high-frequency transmittances we calculate are in agreement with published values.  At Dome~C, the 345~GHz value for an atmosphere with the median PWV of 0.28~mm have been reported to be between 0.85 and 0.90~\citep{Tremblin2012}.  At Dome~A, that same study found 345~GHz transmittance in median conditions to be greater than 0.9.  Both of those findings agree with the spectra we report specifically for March.  At Yangbajing Observatory, for a median atmosphere containing 2.8~mm of water vapor, that same \cite{Tremblin2012} study concluded that YBG has 345~GHz transmittance of 55\%--60\%.  In March, we derive a similar median PWV of 2.4~mm for the YBG site and correspondingly greater transmittance of about 0.67.

\subsection{Improved Sampling of Fourier Components}
\begin{figure*}
\centering
  \includegraphics[width=2.0\columnwidth]{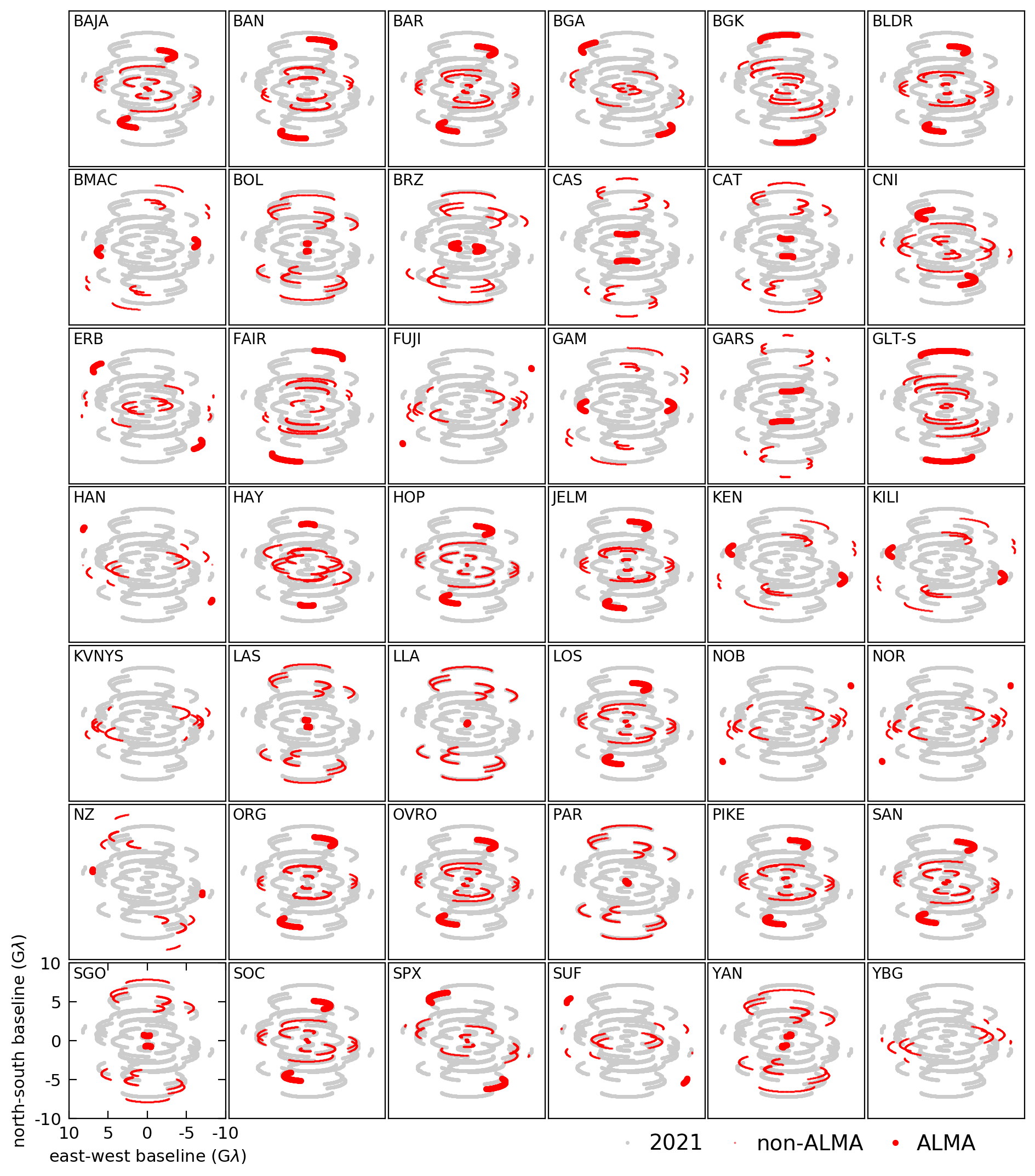}
  \caption{Fourier coverage for \m87 at 230~GHz compared to the 2021 EHT array (gray) for baselines containing the indicated site (red). The large markers denote baselines to ALMA.  Sidebands are omitted for clarity.}
  \label{fig:Fourier_M87}
\end{figure*}

\begin{figure*}
\centering
  \includegraphics[width=2.0\columnwidth]{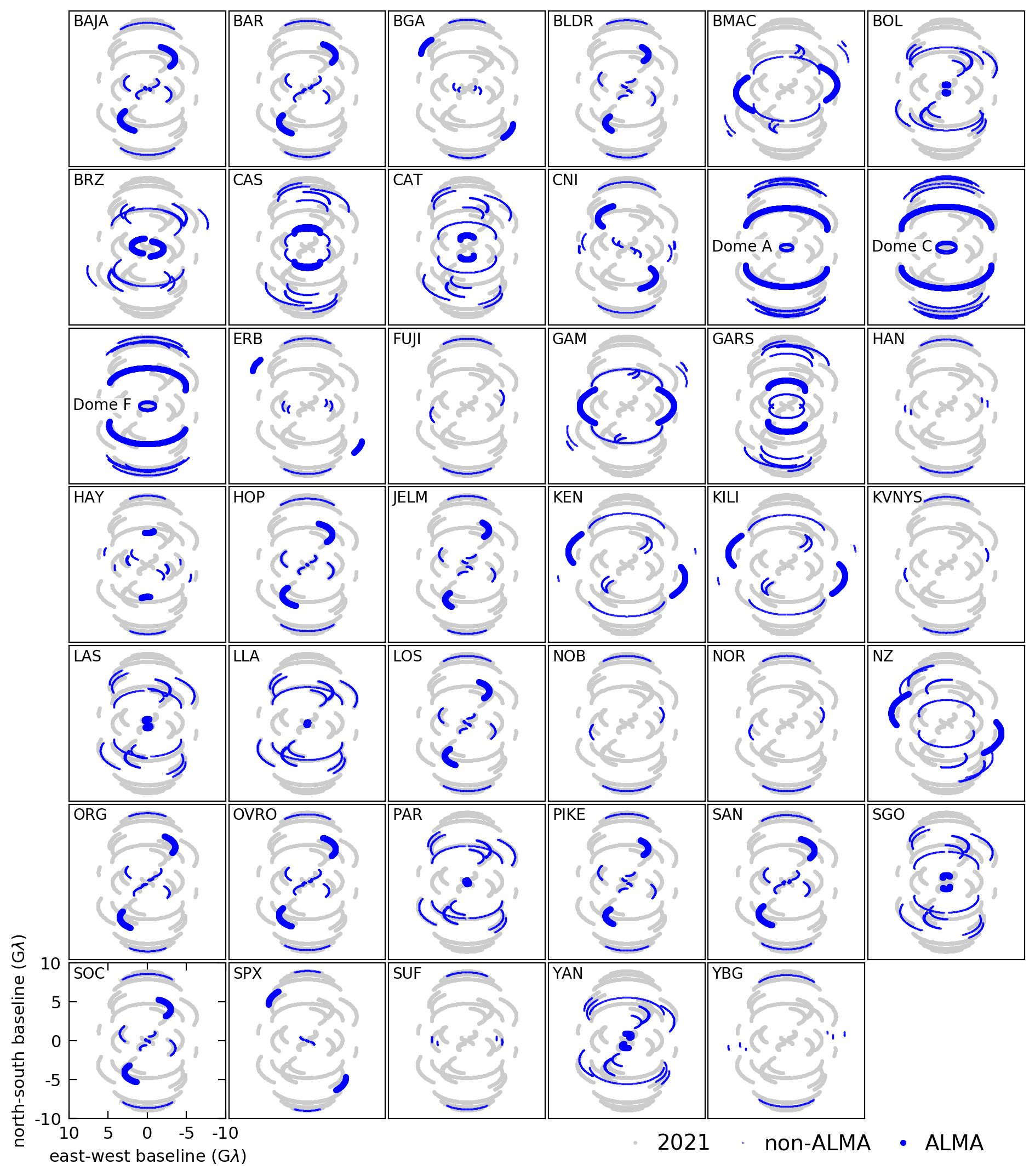}
  \caption{Fourier coverage for \sgra at 230~GHz compared to the 2021 EHT array (gray) for baselines containing the indicated site (blue). The large markers denote baselines to ALMA.  Sidebands are omitted for clarity.}
  \label{fig:Fourier_SgrA}
\end{figure*}

The 230~GHz Fourier coverage for each candidate new site is plotted in Fig.~\ref{fig:Fourier_M87} for \m87 and in Fig.~\ref{fig:Fourier_SgrA} for \sgra.  These coverages are based on the source position and array geometry; we have not flagged low S/N visibilities as we do in the subsequent Monte Carlo filling-factor calculations.  The coverage is for a full 24~hr of observing including times when ALMA cannot observe the source.  Tracks corresponding to baselines with ALMA are plotted in bold.  The tracks of the 2021 EHT array are underlaid in gray for reference.  The 2021 array with ALMA lacks coverage between about 4 and 7~G$\lambda$ in the north--south direction for both \m87 and \sgra.  \sgra also lacks mid-baseline coverage between 1 and 3~G$\lambda$ as well as long-baseline coverage in the east--west direction.  In these plots, an elevation cutoff of 10\degr\ was specified, which is especially important for the very short tracks at large \textit{u}-\textit{v} spacing, e.g., for FUJI and SPX toward \m87.

The number of new baselines is determined by the covisibility of the new site with the 2021 array with ALMA.  For \m87, all of the considered sites that can observe that source form at least five new baselines.  For \sgra, some sites form as few as two new baselines.  The mean longitude of the 2021 EHT sites is 76\degr\ west.  Sites near that longitude have significant covisibility with the existing EHT sites and with ALMA and therefore tend to make long \textit{u}-\textit{v} tracks.

Some site locations are more desirable than others because they sample obvious gaps in Fourier coverage.  For \m87, the BOL, BRZ, CNI, BMAC, GAM, HAY, NZ, and YAN locations fill portions of the 4-7~G$\lambda$ gap in the north--south direction. The BMAC, CAS, CAT, GAM, GARS, NZ, SGO, and YBG sites extend the longest baselines.  For \sgra, the 2021 coverage contains gaps between 4 and 7~G$\lambda$ in the north--south direction, which are filled by CAS, CAT, GARS, KEN, and KILI.  Sites in central and southern Africa, \textit{e.g.}, KILI together with GAM, could significantly improve the resolution of the array in the east--west direction.

\begin{figure*}
\centering
  \includegraphics[width=0.99\linewidth]{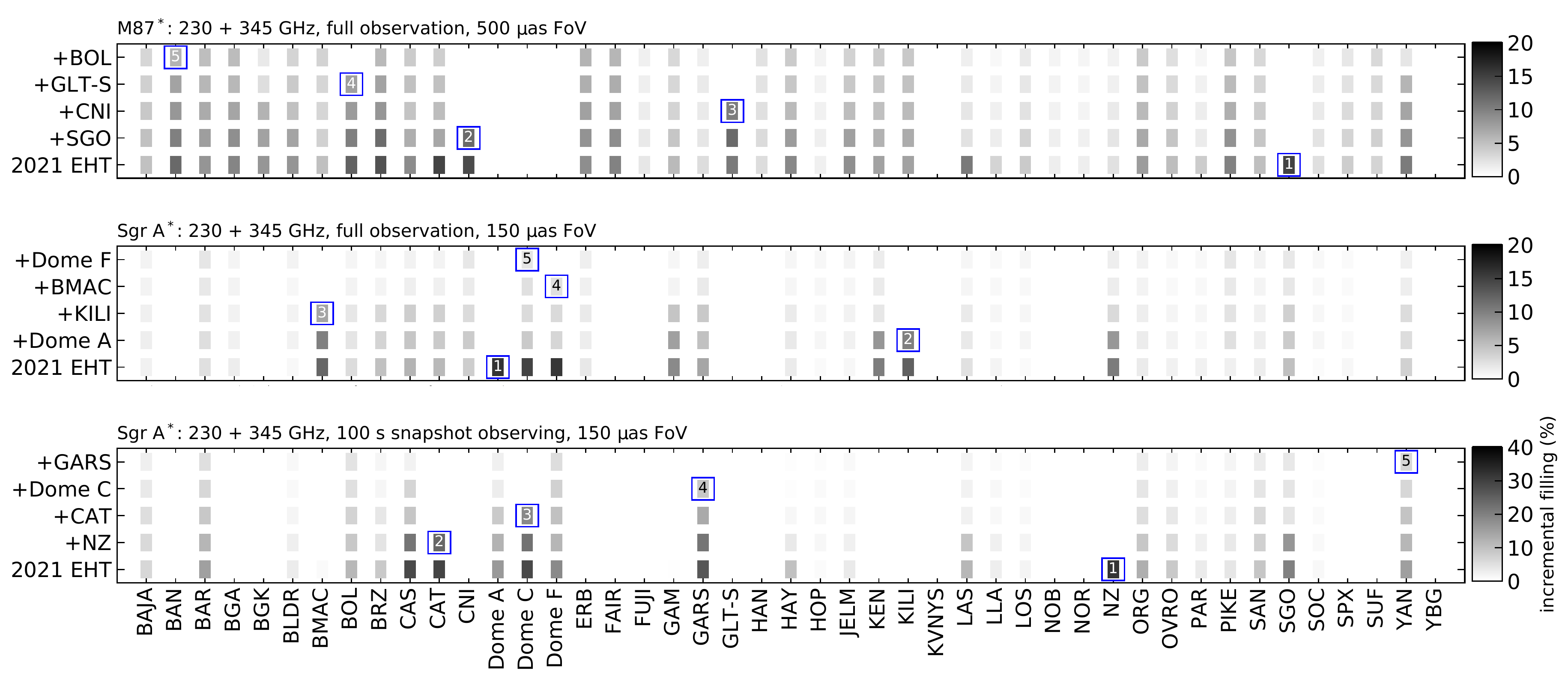}
  \caption{
Incremental filling highlighting the sites with the best performance as the array grows incrementally.  The simulated observations use April probabilistic meteorological conditions, and the filling reported is the mean of the Monte Carlo instantiations.   Beginning with the 2021 EHT array, the incremental filling is calculated for the addition of each new site.  The fiducial array is then augmented by the new site with the greatest incremental filling and the calculation is repeated again.  For \m87 (top panel), the 2021 coverage is improved most by SGO(1) followed by SGO+CNI(2) then by SGO+CNI+GLT-S(3), and so on.  The same procedure was done for \sgra with a full observation (middle), and \sgra with a 100~s snapshot observation (bottom).  The \sgra snapshot observing performance is indicative of the best instantaneous coverage.  The ngEHT observations are assumed to be simultaneous dual-frequency 230 + 345~GHz with 10~m antennas at new stations and an 8~GHz fringe-finding bandwidth in each of two sidebands.  The \m87 fractions are calculated assuming a 500~$\mu$as FoV to encompass the base of the forward jet while the \sgra fractions assume a 150~$\mu$as FoV.  Snapshot fill fractions of zero have no more coverage than the fiducial array.}
  \label{fig:fillfrac}
\end{figure*}

The geometric Fourier coverage contributed by a new site is important, but only if the weather conditions at that new site are adequate for making detections.  The Monte Carlo incremental coverage, which flags intervals with low S/N values based on the probabilistic weather at each site, is presented in Fig.~\ref{fig:fillfrac} for a specific set of observing parameters: 10~m antennas at new stations, simultaneous 230 and 345~GHz frequency coverage, and 8~GHz bandwidth in each of two sidebands.  For the simulations in Figs.~\ref{fig:fillfrac} and \ref{fig:imageM87_Sgra}, the observing schedule is limited to times when ALMA can see the source, which is consistent with previous EHT schedules~\citep{EHT3}.  The bandwidth per frequency band is a projected doubling of the present-day EHT~\citep{EHT2} and helps compensate for some of the additional noise at 345~GHz.  For \m87, an extended 500~$\mu$as FoV is specified, which is motivated by future efforts to connect horizon-scale structure with the \m87 jet~\citep{Blackburn2019a}.  For \sgra, a 150~$\mu$as FoV is used.  We assume an observing time commensurate with the variability timescale of each source.  \m87 evolves slowly, so a full observation is specified to maximize the coverage on each baseline.  \sgra evolves on a timescale of minutes~\citep{EHT2}, so a 100~s snapshot observation is simulated, which truncates the Fourier coverage.  Our approach to flagging low S/N scans accounts for the noise contributed by the physical temperature of the atmosphere.  This is important at low-altitude polar sites like GLT and GARS, where the cold atmosphere compensates for marginal opacity.

The Monte Carlo incremental filling factor is calculated in the following way.  For the full observations, we report the Fourier filling of the augmented and fiducial arrays using the entirety of the Fourier coverage tracks, $F_{\mathrm{aug}}/F_{\mathrm{fid}}-1$.  For the snapshot observations of \sgra, we vary the start time for the fiducial and augmented arrays independently to maximize the filling in each case: $\max{\left(F_{\mathrm{aug}}\right)}/\max{\left(F_{\mathrm{fid}}\right)}-1$.  Depending on the coordinates of a new site, there might be no start times that produce greater snapshot Fourier filling than the fiducial array, which would make the incremental filling for that site zero.  The maximum filling metric is a proxy for the best instantaneous coverage achievable by a new site.

The sites with the greatest incremental filling are added sequentially to the fiducial array as indicated beginning with the 2021 EHT sites.  For \m87, SGO emerges as the site that adds the greatest Fourier coverage to the 2021 array with ALMA.  Repeating the calculation for a 12 station array consisting of the 2021 EHT+SGO, CNI emerges as the next best site for increasing the Fourier coverage.  Similarly for \sgra with a full observation, Dome~A, KEN, BMAC, and Dome~C yield the most incremental coverage. For \sgra with snapshot coverage, NZ adds the most coverage to the 2021 array, followed by Dome~C, then CAT, and GARS.    In general, the incremental coverage for a particular new site decreases as new sites are added to the fiducial array because there are fewer gaps in the Fourier coverage to fill.  Our approach of adding the best locations in sequence makes the problem of selecting a group of sites computationally manageable and leads us to an array with excellent coverage.  As the ngEHT develops, a comprehensive search for the global optimum array could lead to a different selection of sites.
 
The Fourier filling factor we present here is one metric for approaching VLBI array design.  While we favor the filling metric for its simplicity, other considerations and metrics could prioritize a different set of new sites and should be vetted in future design efforts.

\subsection{Imaging with an Expanded Array}
\begin{figure*}
\centering
  \includegraphics[width=0.95\linewidth]{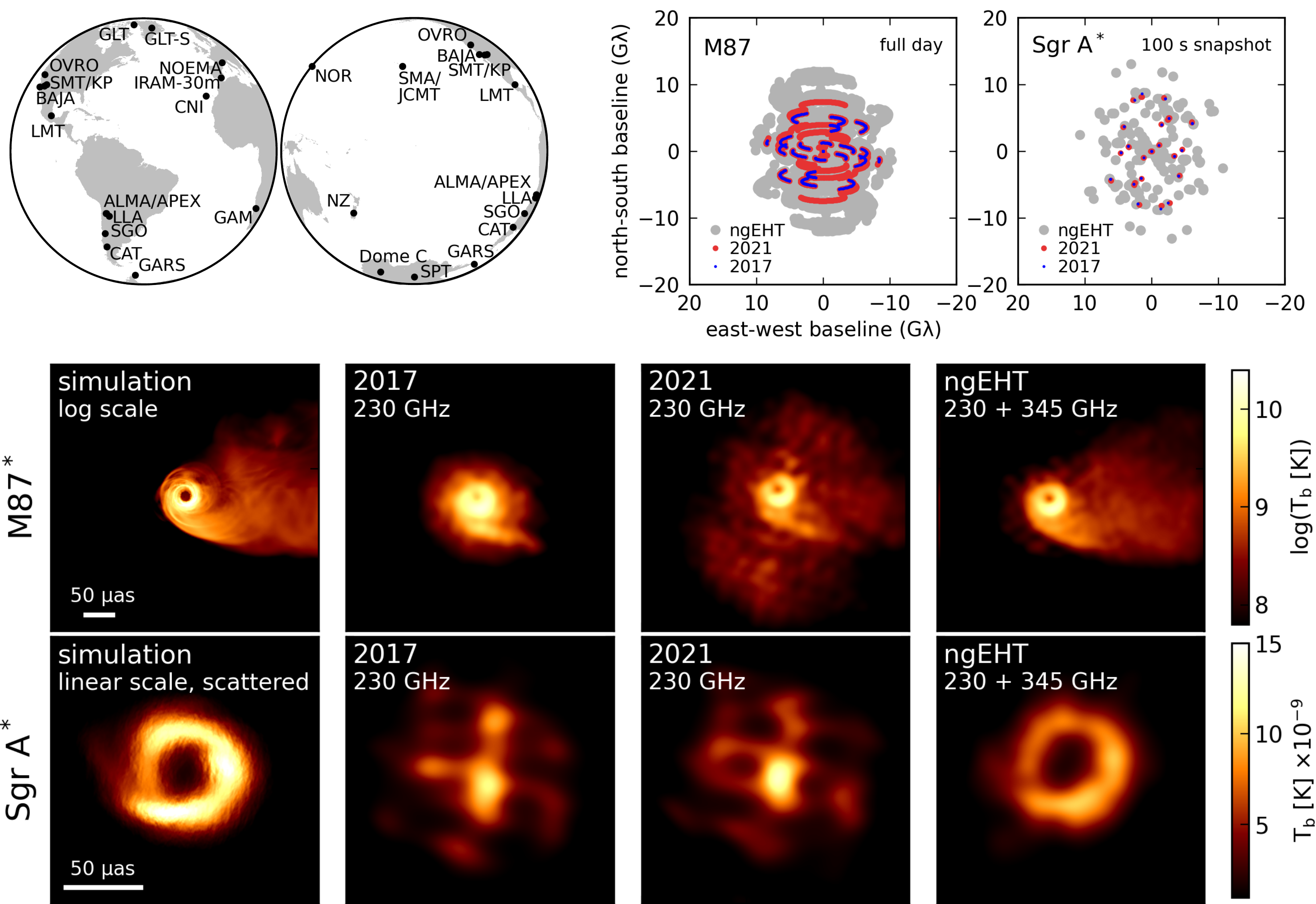}
  \caption{Imaging reconstructions of GRMHD simulations for \m87 (from \cite{Chael2019}, plotted in logarithmic scale flux density) and \sgra (from \cite{Chael2018a}, plotted in linear scale flux density). The simulation panels show the combined 230 and 345~GHz flux distributions where the high-frequency band has been scaled to the lower one using a spectral index obtained from the source model.  We show \sgra as it would appear on the sky, scattered by the interstellar medium~\citep{Johnson2016}.  Using SEFDs calculated in this meteorological study under median conditions for April, the 2017 and 2021 arrays are unable to reconstruct much of the forward-propagating jet in the \m87 model or reconstruct the \sgra brightness distribution with a 100~s snapshot observation.  The ngEHT reconstructions use the array shown on the globe, which augments the 2021 EHT with the leading sites based on the filling factor calculation in Fig.~\ref{fig:fillfrac} plus sites that have good existing infrastructure (BAJA, CAT, CNI, Dome~C, GAM, GARS, GLT-S, LLA, NOR, NZ, OVRO, and SGO).  The ngEHT reconstruction assumes a 10~m antenna diameter for new stations and an 8~GHz bandwidth per sideband.  The 230 and 345~GHz bands are combined in the ngEHT reconstructions using the spectral indices.  The ngEHT dramatically improves the Fourier coverage and reconstructions, both for \m87 and for \sgra in snapshot mode.
}
  \label{fig:imageM87_Sgra}
\end{figure*}

To demonstrate the improvement of horizon-scale imaging that would be possible with an expanded array under realistic meteorological conditions, we compare reconstructions made using the 2017, 2021, and an ngEHT array.  In the 2017 and 2021 arrays, we combine multiple 2~GHz-wide fringe-finding bands for an aggregate bandwidth over two polarizations of 16~GHz in 2017 and 32~GHz in 2021.  For the ngEHT array, we assume the same parameters as the filling factor calculation: dual-frequency 230 and 345~GHz observing with an aggregate bandwidth of 128~GHz over two polarizations.  The bandwidth of the EHT backend has increased at regular intervals~\citep{Doeleman2008,Fish2011,Doeleman2012,Johnson2015,EHT2}, and it is reasonable to anticipate continued improvement on the timescale of array expansion~\citep{Blackburn2019a}.  The image reconstructions are shown in Fig.~\ref{fig:imageM87_Sgra}.  The thermal noise used for these reconstructions are calculated under median conditions.

For \m87, the FoV is extended and the temperature scale is logarithmic to highlight the base of the low-brightness forward jet.  For this particular source model with the 2017 and 2021 arrays, the jet base is not faithfully reconstructed.  In contrast, the additional stations in the ngEHT array and the addition of the 345~GHz band reproduces the jet out to more than five shadow diameters, and the diffuse flux surrounding the jet is more tightly constrained.  The sites in the ngEHT panel were primarily chosen based on the leading sites from the Monte Carlo incremental filling analysis in Fig.~\ref{fig:fillfrac}: three or four sites each for \m87 and snapshot \sgra.  BAJA, GAM, LLA, OVRO, and NOR were also included either because they sample interesting Fourier components or because they have significant infrastructure already in place.  The ngEHT Fourier coverage on \m87 extends to approximately 12~G$\lambda$ under nominal meteorological conditions.

The \sgra reconstruction is shown in linear scale.  We select a start time for the \sgra simulated observation when the Fourier filling is greatest.  With 100~s snapshot observing, the Fourier tracks normally obtained through Earth-aperture synthesis become points.  For the 2017 and 2021 arrays, the Fourier points are sparse, and the arrays and \texttt{eht-imaging} algorithms do not converge on the known structure; however, the large number of baselines in the the ngEHT simulation fills a significant fraction of the Fourier plane even in snapshot mode.  The ngEHT is capable of reconstructing the ring emission of \sgra including the morphology of the discontinuity in the left-hand site of the image.  A full multifrequency reconstruction with scattering mitigation is an active research topic; however, these short-timescale reconstructions suggest that detailed movie-making of dynamical features at \sgra could be possible with the ngEHT.

\section{Conclusions}
We have evaluated potential new sites for the ngEHT and examined how they improve the capability of the EHT array to study \m87 and \sgra in increasing detail. We have used meteorological statistics derived from the MERRA-2 reanalysis to develop a list of 45 potential new sites.  Our site water vapor reanalyses of meteorological data are in close agreement with available in-situ measurements.  We find that a number of the candidate sites have, first, good conditions for submillimeter observations and, second, also sample new spatial frequencies on both large and small angular scales, which significantly enhances imaging fidelity of the EHT array.  The Monte Carlo incremental filling calculation suggests that the priority sites strongly depend on details like the source and the FoV.  Future work is planned for assessing the best array across a range of science cases and metrics; however, our example imaging reconstruction shows that under realistic median meteorological conditions, additional stations in the ngEHT will improve the ability to image low-flux, dynamical features for \m87 and \sgra.  Future work will also include field measurements over at least a full seasonal cycle to validate MERRA-2 against local conditions.

\acknowledgments
We thank several members of the EHT Collaboration and T. K. Sridharan for suggesting potential new sites for study, Simon Radford for useful discussions about the SMA opacity logs, Andrew Chael for providing GRMHD source models, Antonio Fuentes whose MPI parallelization framework we adapted for the fill fraction calculations, Dom Pesce and Greg Lindahl for their assistance with the filling fraction code, and Sara Issaoun, Mareki Honma, Lynn Matthews, and Peter Galison for their useful feedback on the manuscript.  We acknowledge the significance that Maunakea, where the SMA and JCMT EHT stations are located, has for the Hawaiian people.  This work was supported in part by the National Science Foundation (AST-1935980, AST-1716536, OISE-1743747), the Gordon and Betty Moore Foundation (GBMF-5278), and the Black Hole Initiative at Harvard University, through a grant (60477) from the John Templeton Foundation.

\bibliographystyle{aasjournal}
\bibliography{references}

\end{document}